\documentclass{aa} % [referee]
\usepackage{ae, aecompl, amsmath, amssymb, bm, booktabs, graphicx, hyperref, longtable, lscape, newtxtext, subcaption, textcomp, txfonts, xcolor}
\usepackage[T1]{fontenc}
\usepackage[separate-uncertainty = true, multi-part-units = single]{siunitx}
\newcommand\bred[1]{{{#1}}}
\begin{document}

\title{Blazars in the LOFAR Two-Metre Sky Survey First Data Release}
\author{S. Mooney\inst{1}\fnmsep\thanks{sean.mooney@ucdconnect.ie}
\and J. Quinn\inst{1}
\and J. R. Callingham\inst{2}
\and R. Morganti\inst{2, 3}
\and K. Duncan\inst{4}
\and L. K. Morabito\inst{5}
\and P. N. Best\inst{6}
\and G. G\"urkan\inst{7}
\and M. J. Hardcastle\inst{8}
\and I. Prandoni\inst{9}
\and H. J. A. R\"ottgering\inst{4}
\and J. Sabater\inst{6}
\and T. W. Shimwell\inst{2}
\and A. Shulevski\inst{10}
\and C. Tasse\inst{11, 12}
\and W. L. Williams\inst{8}}
\institute{School of Physics, University College Dublin, Belfield, Dublin 4, Republic of Ireland
\and ASTRON, Netherlands Institute for Radio Astronomy, PostBus 2, 7990 AA, Dwingeloo, The Netherlands
\and Kapteyn Astronomical Institute, University of Groningen, P.O. Box 800, 9700 AV Groningen, The Netherlands
\and Leiden Observatory, Leiden University, PO Box 9513, NL-2300 RA Leiden, the Netherlands
\and Astrophysics, University of Oxford, Denys Wilkinson Building, Keble Road, Oxford OX1 3RH, UK
\and SUPA, Institute for Astronomy, Royal Observatory, Blackford Hill, Edinburgh, EH9 3HJ, UK
\and CSIRO Astronomy and Space Science, PO Box 1130, Bentley WA 6102, Australia
\and Centre for Astrophysics Research, School of Physics, Astronomy and Mathematics, University of Hertfordshire, College Lane, Hatfield AL10 9AB, UK
\and INAF - Istituto di Radioastronomia, Via P. Gobetti 101, 40129 Bologna, Italy
\and Anton Pannekoek Institute for Astronomy, University of Amsterdam, Postbus 94249, 1090 GE Amsterdam, The Netherlands
\and GEPI \& USN, Observatoire de Paris, Universit\'e PSL, CNRS, 5 Place Jules Janssen, 92190 Meudon, France
\and Department of Physics \& Electronics, Rhodes University, PO Box 94, Grahamstown, 6140, South Africa}
\date{Received 24 July 2018; accepted 18 October 2018}
\abstract{Historically, the blazar population has been poorly understood at low frequencies because survey sensitivity and angular resolution limitations have made it difficult to identify megahertz counterparts. We used the LOFAR Two-Metre Sky Survey (LoTSS) first data release value-added catalogue (LDR1) to study blazars in the low-frequency regime with unprecedented sensitivity and resolution. We identified radio counterparts to all $98$ known sources from the Third \textit{Fermi}-LAT Point Source Catalogue (3FGL) or Roma-BZCAT Multi-frequency Catalogue of Blazars ($5^{\mathrm{th}}$ edition) that fall within the LDR1 footprint. Only the 3FGL unidentified $\gamma$-ray sources (UGS) could not be firmly associated with an LDR1 source; this was due to source confusion. We examined the redshift and radio luminosity distributions of our sample, finding flat-spectrum radio quasars (FSRQs) to be more distant and more luminous than BL Lacertae objects (BL Lacs) on average. Blazars are known to have flat spectra in the gigahertz regime but we found this to extend down to \SI{144}{\mega\hertz}, where the radio spectral index, $\alpha$, of our sample is $-0.17 \pm 0.14$. For  BL Lacs, $\alpha = -0.13 \pm 0.16$ and for FSRQs, $\alpha = -0.15 \pm 0.17$. We also investigated the radio-to-$\gamma$-ray connection for the $30$ $\gamma$-ray-detected sources in our sample. We find Pearson's correlation coefficient is $0.45$ ($p = 0.069$). This tentative correlation and the flatness of the spectral index suggest that the beamed core emission contributes to the low-frequency flux density. We compare our sample distribution with that of the full LDR1 on colour-colour diagrams, and we use this information to identify possible radio counterparts to two of the four UGS within the LDR1 field. We will refine our results as LoTSS continues.}
\maketitle

\section{Introduction} \label{sec:introduction}

The centres of some galaxies are extremely luminous, producing broadband non-thermal emission. These compact regions are known as active galactic nuclei (AGN). Some fraction of AGN are understood to have relativistic jets and by chance some of the jets are orientated close to our line of sight. Such AGN are known as blazars \citep[see the review by][]{1995PASP..107..803U}. The jets are believed to be powered by the accretion of matter onto supermassive black holes residing at the galactic cores. Relativistic beaming effects give rise to apparent superluminal motion, and Doppler boosting increases the observed luminosity. Although blazars are the most common sources in the $\gamma$-ray regime \citep{2015ApJS..218...23A}, only a small number of blazars are $\gamma$-ray-loud and the reasons for this are still unclear \citep{2012RAA....12.1475F}. % However, in recent decades, blazar science has been rejuvenated. This has been, in part, due to the current generation of telescopes probing these high-energy regimes, such as the Fermi $\gamma$-ray Space Telescope (\textit{Fermi}). For example, \cite{2015A&ARv..24....2M} outline how \textit{Fermi} results have enabled detailed modeling of the particle acceleration processes that produce the $\gamma$-rays.

There are two types of blazars that are distinguished by their observational properties: BL Lacertae objects (BL Lacs) and flat-spectrum radio quasars (FSRQs). The populations are defined by the presence or absence of strong emission lines, which is controlled by the inner accretion disc. BL Lacs possess featureless optical spectra and are generally associated with beamed jet-mode (radiatively inefficient) AGN. In contrast, strong optical emission lines are a characteristic of FSRQs and they are often associated with beamed radiative-mode AGN. However, one commonality shared by BL Lacs and FSRQs is the broadband nature of the radiation they emit. % , and a high fraction of this luminosity resides in hotspots in the lobes.

% FSRQs are often observed to be more distant and more luminous than BL Lacs.

% \bred{, where} the brightness of falls off with distance from the centre of the galaxy \citep{1995PASP..107..803U}

% While the majority of Fanaroff-Riley type I and type II sources are associated with jet-mode and radiative-mode AGN respectively, the correspondence is not perfect \citep{2014ARA&A..52..589H}. 

The characteristic structure seen in the spectral energy distributions (SEDs) for blazars consists of two components in the $\nu F_{\nu}$--$\nu$ plane (where $\nu$ is frequency and $F_\nu$ is flux), which has the non-thermal components dominating energetically over the thermal component at all wavelengths. These two components give the blazar SEDs their characteristic double-humped shape.

The first component begins in the radio waveband and peaks in the optical or X-ray waveband. This emission can be attributed to synchrotron processes from a population of relativistic ($\gtrsim \si{keV}$) electrons in a magnetic field. Blazars typically possess flat spectra at gigahertz frequencies, where the radio spectral index, $\alpha$, is defined as $S(\nu) \propto \nu^{\alpha}$, typically $\alpha > -0.5$. \cite{2014ApJS..212....3N} found that blazars have flat spectra down to ${\sim}\SI{300}{\mega\hertz}$. At lower frequencies, the spectrum becomes inverted (i.e. $\alpha > 0$) because of synchrotron self-absorption.

\begin{table*}
\centering
\caption{Breakdown of our sample is shown according to catalogue and source type. The 3FGL includes 3LAC; the sole difference between these catalogues over the LDR1 footprint is the four UGS that are included in 3FGL only.}
\label{tab:1}
\begin{tabular*}{\linewidth}{@{\extracolsep{\fill}}lcccccccccc@{}}
\toprule
\;\;\textbf{Catalogue}      & \textbf{FSRQ} & \textbf{BL Lac} & \textbf{\begin{tabular}[c]{@{}c@{}}Uncertain\\ type\end{tabular}} & \textbf{\begin{tabular}[c]{@{}c@{}}BL Lac\\ candidate\end{tabular}} & \textbf{\begin{tabular}[c]{@{}c@{}}Blazar\\ candidate\end{tabular}}& \textbf{\begin{tabular}[c]{@{}c@{}}Galaxy\\ dominated\end{tabular}} & \textbf{\begin{tabular}[c]{@{}c@{}}Radio\\ galaxy\end{tabular}} & \textbf{UGS} & \textbf{Total} \\ \midrule
\;\;\textbf{BZCAT only}     & 41            & 12              & 4                       & 1                       &0  & 10                        & 0            & 0                       & \textbf{\;\,68}    \\
\;\;\textbf{3FGL only}      &  \;\,0            &  \;\,0              & 0                       & 0&3                         &  \;\,0                                   & 1                     & 4            & \textbf{\;\,\,\;\,8}     \\
\;\;\textbf{BZCAT and 3FGL} &  \;\,8            & 15              & 3                       & 0                      &0   &  \;\,0                        & 0            & 0                               & \textbf{\;\,26}    \\
\;\;\textbf{Total}          & \textbf{49}   & \textbf{27}     & \textbf{7}              & \textbf{1}  &\textbf{3}              & \textbf{10}               & \textbf{1}            & \textbf{4}   & \textbf{102}   \\ \bottomrule
\end{tabular*}%
\end{table*}

The second feature of the SED peaks between the \si{\mega\electronvolt} and \si{\tera\electronvolt} energy bands and may be caused by inverse-Compton scattering \citep[e.g.][]{1994ApJ...421..153S} but this remains an open question \citep{2012int..workE..69B}. If this is the case, then seed photons originating from the synchrotron process are inverse-Compton scattered by the electrons in the jet to higher energies \citep[i.e. synchrotron self-Compton radiation;][]{1985ApJ...298..114M}. However, it is also possible that the seed photons originate from outside the jet -- for example, from the accretion disc or broad line region. Alternatively, the high-energy peak of the SED may be the result of hadronic synchrotron processes, rather than leptonic inverse-Compton processes \citep{Bottcher2007}.

We search for a correlation between the low-frequency radio emission and the $\gamma$-ray emission in this study. The existence of such a correlation is still debated \citep{2012ApJ...751..149P}. Several studies have found a correlation \citep{1993ApJ...410L..71S, 1993MNRAS.260L..21P, 1994ApJ...430L..21S, 2011ApJ...741...30A, 2012ApJ...757...25L}. However, taking all biases into account, such as the limited dynamic range (when considering flux densities) or the common redshift dependence (when considering luminosities) is non-trivial \citep{2009ApJ...696L..17K}. For example, \cite{1997A&A...320...33M} and \cite{1998ApJ...496..752C} disputed evidence of a correlation on the grounds of redshift biases and the sensitivity limits of the surveys used. % Conclusive proof of a correlation between the flux density at self-absorbed radio frequencies (${\sim}\SI{100}{\mega\hertz}$) and at $\gamma$-ray frequencies (${\sim}\SI{100}{\mega\electronvolt}$) would suggest that the non-thermal processes which produce the two components on the SED are intrinsically related to each other, and evolve in a similar manner. Should the radio and $\gamma$-ray emission be uncorrelated, it would suggest that the two emission regions operate independently.

\bred{Studying blazars} at megahertz frequencies \bred{is challenging} because their characteristic flat spectra make it difficult to identify counterparts in this regime. For example, \cite{2016A&A...588A.141G} used the Murchison Widefield Array Commissioning Survey (MWACS) \citep{2014PASA...31...45H} to examine the 120--\SI{180}{\mega\hertz} emission from blazars. The MWACS has $\sim$\SI{3}{\arcminute} angular resolution and a typical noise level of \SI{40}{mJy \, beam^{-1}}, which allowed \cite{2016A&A...588A.141G} to identify low-frequency counterparts to $186$ of $517$ ($36\%$) blazars in the MWACS footprint. \cite{2016A&A...588A.141G} then calculated the mean low-frequency spectral index to be $-0.57 \pm 0.02$, and identified a mild correlation between the radio flux density and the $\gamma$-ray energy flux ($r = 0.29$, $p = 0.061$). \cite{2017ApJ...836..174C} also identified a small number of blazars that show a peaked spectrum in the low-frequency spectra from the GLEAM survey \citep{2017MNRAS.464.1146H}. This emphasises that a simple selection of flat-spectrum radio sources may not select all blazars. However, both \cite{2017ApJ...836..174C} and \cite{2016A&A...588A.141G} were limited in their resolution and sensitivity to explore the population in depth.

We use the LOFAR Two-Metre Sky Survey (LoTSS) first data release value-added catalogue (LDR1) to study the \SI{144}{\mega\hertz} properties of blazars \citep{2018A&A...,wendy,ken}. We cross-matched LDR1 with the Third \textit{Fermi}-LAT Point Source Catalogue (3FGL) \citep{2015ApJS..218...23A}, the Roma-BZCAT Multi-frequency Catalogue of Blazars ($5^{\mathrm{th}}$ edition) \citep{2015Ap&SS.357...75M}, and the very-high-energy catalogue called TeVCAT \citep{2008ICRC....3.1341W}. The LDR1 catalogue covers \SI{424}{deg^{2}} of the sky with future data releases aiming to significantly expand this to full coverage of the northern sky. In this respect, this work paves the way for a larger study with future data releases.%\LEt{Please avoid beginning sentences with abbreviations, acronyms, numbers in figures, and the like. Please check for this throughout the paper.  See Sect. 6.2.1 (http://www.aanda.org/author-information/language-editing/1-introduction).}

This paper is organised as follows: The sample of sources used for this study was constructed from several surveys and catalogues, each of which is described in turn in \S\ref{sec:sac}. The way in which we built our sample is detailed in \S\ref{sec:analysis}. Our results are presented in \S\ref{sec:results} and discussed in \S\ref{sec:conclusions}. We use a $\Lambda$CDM cosmological model throughout this paper with $h = 0.71$, $\Omega_{\mathrm{m}} = 0.26$, and $\Omega_{\Lambda} = 0.74,$ where $H_{0} = 100 h \, \si{\kilo \metre \, \second^{-1} \, Mpc^{-1}}$ is the Hubble constant. We maintain the definition of $\alpha,$ where $S(\nu) \propto \nu^{\alpha}$.

\section{Surveys and catalogues} \label{sec:sac}

\subsection{LOFAR Two-Metre Sky Survey First Data Release} \label{sbs:lt-mss}

The LOw Frequency ARray (LOFAR) is a radio interferometer with stations located throughout Europe \citep{2013A&A...556A...2V}. The LOFAR Surveys key science project aims to map the sky above the northern hemisphere between \SI{120}{\mega\hertz} and \SI{168}{\mega\hertz}. A full description of the LoTSS can be found in \cite{2017A&A...598A.104S}.  The LoTSS is underway, making use of the core and remote LOFAR stations in the Netherlands.

The first data release is outlined in \bred{\cite{2018A&A...}} and the value-added catalogue is outlined in \bred{\cite{wendy}} and \bred{\cite{ken}}. The LDR1 uses data collected between $2014$ May $23$ and $2015$ October $15$, focussing on the HETDEX Spring Field \citep{2008ASPC..399..115H}. The right ascension ranges approximately from \SI{10}{\hour} \SI{45}{\m}~0\SI{0}{\second} to \SI{15}{\hour}~\SI{30}{\m}~0\SI{0}{\second} and the declination ranges approximately from \SI{45}{\degree}~0\SI{0}{\arcminute}~0\SI{0}{\arcsecond} to \SI{57}{\degree}~0\SI{0}{\arcminute}~0\SI{0}{\arcsecond}; the advantage of this region for the study of blazars is that it is far from the galactic centre. Furthermore, the \SI{6}{\arcsecond} angular resolution and \SI{71}{\micro Jy \, beam^{-1}} median sensitivity of LDR1 is unrivaled with respect to existing radio surveys. \bred{To study the blazar population, we use this catalogue described in \cite{2018A&A...}, which has the direction-dependent corrections applied.}

\subsection{3FGL and 3LAC} \label{sbs:3fgl}

% The Large Area Telescope (LAT) is a $\gamma$-ray detector on board \textit{Fermi}. The 3FGL is based on scientific data from the first four years of \textit{Fermi}-LAT, covering the 0.1--\SI{300}{\giga\electronvolt} energy range \citep{2015ApJS..218...23A}. There are $3\,033$ sources in 3FGL. Of these, $498$ ($16\%$) are FSRQs, $669$ ($22\%$) are BL Lacs, $1\,009$ ($33\%$) are unassociated $\gamma$-ray sources (UGS), and the remaining $857$ ($28\%$) are other types of sources. The UGS are the detections to which a known source could not be unambiguously linked, often due to source confusion.

The 3FGL is based on data from the first four years of \textit{Fermi}-LAT, covering the 0.1--\SI{300}{\giga\electronvolt} energy range \citep{2015ApJS..218...23A}. \bred{There are $3\,033$ sources in 3FGL, of which $1\,009$ are} unassociated $\gamma$-ray sources (UGS). These are sources to which a known source could not be unambiguously linked, often due to source confusion.

The Third Catalogue of AGN detected by \textit{Fermi}-LAT (3LAC) is the most comprehensive catalogue of $\gamma$-ray AGN at present. The 3LAC is based on 3FGL sources that have a test statistic $>25$ (i.e. $\gtrsim5 \sigma$ significance) between \SI{100}{\mega\electronvolt} and \SI{300}{\giga\electronvolt} over the period extending from 2008 August 04 to 2012 July 31 \citep{2015ApJ...810...14A}. The 3LAC contains $1\,773$ AGN in total with $491$ ($28\%$) FSRQs, $662$ ($37\%$) BL Lacs, $585$ ($33\%$) blazars of unknown type, and $35$ ($2\%$) sources of other types. We use the improved source positions and blazar classification information in 3LAC to aid in the study of our sample. %The 3LAC sources have been matched with known AGN by means of likelihood ratio \citep{2015ApJ...810...14A} or Bayesian statistics \citep{2015ApJS..218...23A}. 

\subsection{BZCAT} \label{sbs:bzcat}

BZCAT is a catalogue of blazars that contains multi-frequency data from a number of surveys \citep{2015Ap&SS.357...75M}. The BZCAT contains radio flux measurements which are either at \SI{1.4}{\giga\hertz} from the National Radio Astronomy Observatory Very Large Array Sky Survey (NVSS) \citep{1998AJ....115.1693C} (\SI{0.45}{mJy\,beam^{-1}} sensitivity) or at \SI{0.8}{\giga\hertz} from the Sydney University Molonglo Sky Survey (SUMSS). For the region of sky we are interested in, the radio flux measurements used are those at \SI{1.4}{\giga\hertz} because sources in LDR1 have a declination $>\SI{-30}{\degree}$. The source positions are mostly derived from very-long-baseline interferometry measurements. In addition, BZCAT reports information from the Wide-field Infrared Survey Explorer (\textit{WISE}) and the Sloan Digital Sky Survey (SDSS) optical database.

Edition 5.0.0 of BZCAT was used \bred{and, while 3FGL contains a higher fraction of BL Lacs, BZCAT lists mostly FSRQs. Of the ${3\,561}$ sources in BZCAT,} $1\,909$ ($54\%$) are FSRQs, $1\,425$ ($40\%$) are BL Lacs, $227$ ($6\%$) are blazars of uncertain type, and $274$ ($8\%$) are galaxy-dominated blazars. 

\subsection{TeVCAT} \label{sbs:othersurveys}

We searched for sources in TeVCAT \citep{2008ICRC....3.1341W}, which provides \si{\tera\electronvolt} data, but found no sources within the LDR1 footprint. However, this will become an important source of information with which to study blazars as LoTSS progresses.

\section{Analysis} \label{sec:analysis}

\subsection{Sample construction}

For the high-energy sources, BZCAT positional data were used where available, and 3LAC data were used secondarily. Both have accurately defined positions. Likewise, the source classifications (FSRQs, BL Lacs, etc.) were taken from BZCAT in the first instance and from 3LAC for sources without a BZCAT association. For the UGS, 3FGL positions were used, which had comparatively large uncertainty ellipses.

Using \textsc{Topcat} \citep[Tool for Operations on Catalogues and Tables;][]{2005ASPC..347...29T}, we cross-matched the catalogues with LDR1, \bred{where the LDR1 positions take account of any extended features, not just the core regions.} We implemented a \SI{12}{\arcsecond} search radius. Although this is comparatively large compared to the astrometric uncertainties (the average uncertainty on the position of an LDR1 source is ${\sim}\SI{0.3}{\arcsecond}$), $93\%$ of LDR1 sources were unique matches within \SI{7}{\arcsecond} of the BZCAT or 3LAC positions. The seven sources with a separation of 7--\SI{12}{\arcsecond} \bred{are extended in LDR1, and also had unique matches within \SI{12}{\arcsecond}.} All matches were confirmed visually and images showing the sources in our sample along with the BZCAT/3LAC positions can be found at \url{https://github.com/mooneyse/LDR1-blazars}. % \bred{There are only $13$ compact single Gaussian sources, and $85$ sources to which multiple Gaussians were fit.}

 %Fig.~\ref{fig:11-41}.}

An overview of the $102$ unique extragalactic sources in our sample is given in Table~\ref{tab:1}, where $68$ sources were in BZCAT only and therefore have no $\gamma$-ray detection. An LDR1 match was found for all sources, excluding the UGS; a unique match could not be determined for the four UGS because of source confusion.

To investigate the likelihood of spurious detections, we shifted all sources in BZCAT by \SI{2}{\degree} in a random direction and performed the same cross-matching procedure as before. We repeated this several times and found no matches within $\SI{7}{\arcsecond}$ and ${\sim}2$ matches within $\SI{10}{\arcsecond}$, indicating that it is likely our sample is free from such spurious detections.

\subsection{Radio spectral index} \label{sbs:rsi}

To calculate the radio spectral indices, flux density measurements from several surveys in the 0.07--\SI{1.4}{\giga\hertz} range were employed: the Very Large Array Low-frequency Sky Survey Redux (VLSSr) \citep{2014MNRAS.440..327L}, the 7\textsuperscript{th} Cambridge Survey of Radio Sources (7C), the Westerbork Northern Sky Survey (WENSS) \citep{1997A&AS..124..259R}, and NVSS were used where available. Table~\ref{tab:surveys} shows the frequency corresponding to each survey and the number of sources for which each survey had data. \bred{The TIFR Giant Metrewave Radio Telescope Sky Survey First Alternative Data Release (TGSS ADR1) \citep{2017A&A...598A..78I} was not used in spectral fitting but is shown in Table~\ref{tab:surveys} to allow for comparisons.}

\begin{table}
\centering
\caption{Number of sources found in each catalogue or survey. NVSS detected $97$ of $98$ BZCAT or 3FGL sources in the field. There are $8$ sources detected in LDR1 and NVSS only.}
\label{tab:surveys}
\begin{tabular*}{\columnwidth}{@{\extracolsep{\fill}}lccccc@{}}
\toprule
\multicolumn{1}{l}{\textbf{Survey}} & \textbf{\begin{tabular}[c]{@{}c@{}}$\bm{\nu}$\\ (MHz)\end{tabular}} & \textbf{BL Lacs} & \textbf{FSRQs} & \textbf{Other} & \multicolumn{1}{l}{\textbf{Total}} \\ \midrule
\;VLSSr &  \,\;\(73.8\) & $\,\;5$ & $24$ & $\,\;7$ & \textbf{36} \\
\;LDR1 & \,\,\;\(144\) & $27$ & $49$ & $22$ & \textbf{98} \\
\;TGSS & \,\;\,\(148\) & $20$ & $45$ & $17$ & \textbf{82} \\
\;7C & \,\;\,\(151\) & $10$ & $37$ & $13$ & \textbf{60} \\
\;WENSS & \,\,\;\(325\) & $23$ & $47$ & $19$ & \textbf{89} \\
\;NVSS & \(1\,400\) & $26$ & $49$ & $22$ & \textbf{97} \\ \bottomrule
\end{tabular*}
\end{table}

\begin{table*}
\centering
\caption{Summary of our results is shown. The flux density ($S_{144\,\mathrm{MHz}}$) and luminosity ($L_\nu$) refer to median values; for the redshift ($z$) and spectral index ($\alpha$) the average is given. The number of $\gamma$-ray-detected sources is $N_\gamma$.}
\label{tab:loudness}
\begin{tabular*}{\linewidth}{@{\extracolsep{\fill}}lcccccc@{}}
\toprule
\;\;\textbf{Subsample}  & \textbf{\textit{N}} & \textit{\textbf{z}}& \textbf{\begin{tabular}[c]{@{}c@{}}\textit{S}\textsubscript{144\,MHz}\\ (mJy)\end{tabular}} & \textbf{\begin{tabular}[c]{@{}c@{}}$\bm{L_\nu}$\\ (W Hz)\end{tabular}} &$\bm{ \alpha }$& \textbf{\textit{N}}$\bm{_\gamma}$ \;\;      \\ \midrule
\;\;BL Lacs        & $27$         & 0.776          & \,\;$69\pm 14$        & $\num{3.7\pm0.7e25}$                                        & $-0.13 \pm 0.16$           &  15 \;\;        \\
\;\;FSRQs          & $49$         & 1.249          & $362 \pm 72$          & $\num{1.4\pm0.3e27}$                                        & $-0.15 \pm 0.16$           & \,\;8  \;\;     \\
\;\;Others         & $22$         & 0.466          &  $152 \pm 30$         & $\num{5.8\pm1.2e25}$                                        & $-0.28\pm 0.19$            & \,\;7    \;\;   \\
\;\;\textbf{Total} &\textbf{ 98 } & \textbf{0.947} & \textbf{199 $\pm$ 40} & \textbf{(2.7 $\pm$ 0.5) ${\times }$ 10\textsuperscript{26}} & \textbf{--0.17 $\pm$ 0.14} & \textbf{30} \;\; \\ \bottomrule
\end{tabular*}
\end{table*}

% A simple power-law was fit to the data to find $\alpha$, where the appropriate errors were taken into account using the spectral fitting method outlined in \cite{2015ApJ...809..168C}.

\bred{The spectral modelling performed was identical to that done by \cite{2015ApJ...809..168C}. In summary, the  Markov chain Monte Carlo (MCMC) algorithm \texttt{emcee} \citep{2013PASP..125..306F} was used to sample the posterior probability density functions of a power-law or a curved-power-law model \citep[see equations 1 and 2 of][]{2017ApJ...836..174C}. Physically sensible priors were applied (such as that the normalisation constant cannot be negative) and a Gaussian likelihood function was maximised by applying the simplex algorithm to direct the walkers \citep{doi:10.1093/comjnl/7.4.308}. For this method, the uncertainties reported on the flux density values in all the surveys were assumed to be Gaussian and independent.} 

We compared the modelled spectral index to the spectral index where only the lowest \bred{(VLSSr where available, but LDR1 in the majority of cases)} and highest frequencies \bred{(NVSS)} were used, $\alpha_{\mathrm{min-max}}$. We found $\alpha_{\mathrm{min-max}} = {{-0.20}} \pm 0.14$, and this is in agreement with $\alpha = -0.17 \pm 0.14$, when all points are used. The equation of the line between these quantities is $\alpha_{\mathrm{min-max}} = 0.97 \alpha - 0.03$, where $r = 0.96$, indicating that fitting a power law to the sources in our sample is a valid assumption.

\bred{The majority ($83\%$) of our sources are found in both TGSS and NVSS and hence appear in the TGSS-to-NVSS spectral index catalogue \citep{2018MNRAS.474.5008D}. These sources have an average TGSS-to-NVSS spectral index of $-0.28 \pm 0.15$, which is in keeping with our result of ${\alpha = -0.24 \pm 0.14}$ for the same sample.}

\begin{figure*}
    \centering
    \begin{subfigure}{0.33\textwidth}
                \includegraphics[width=\textwidth]{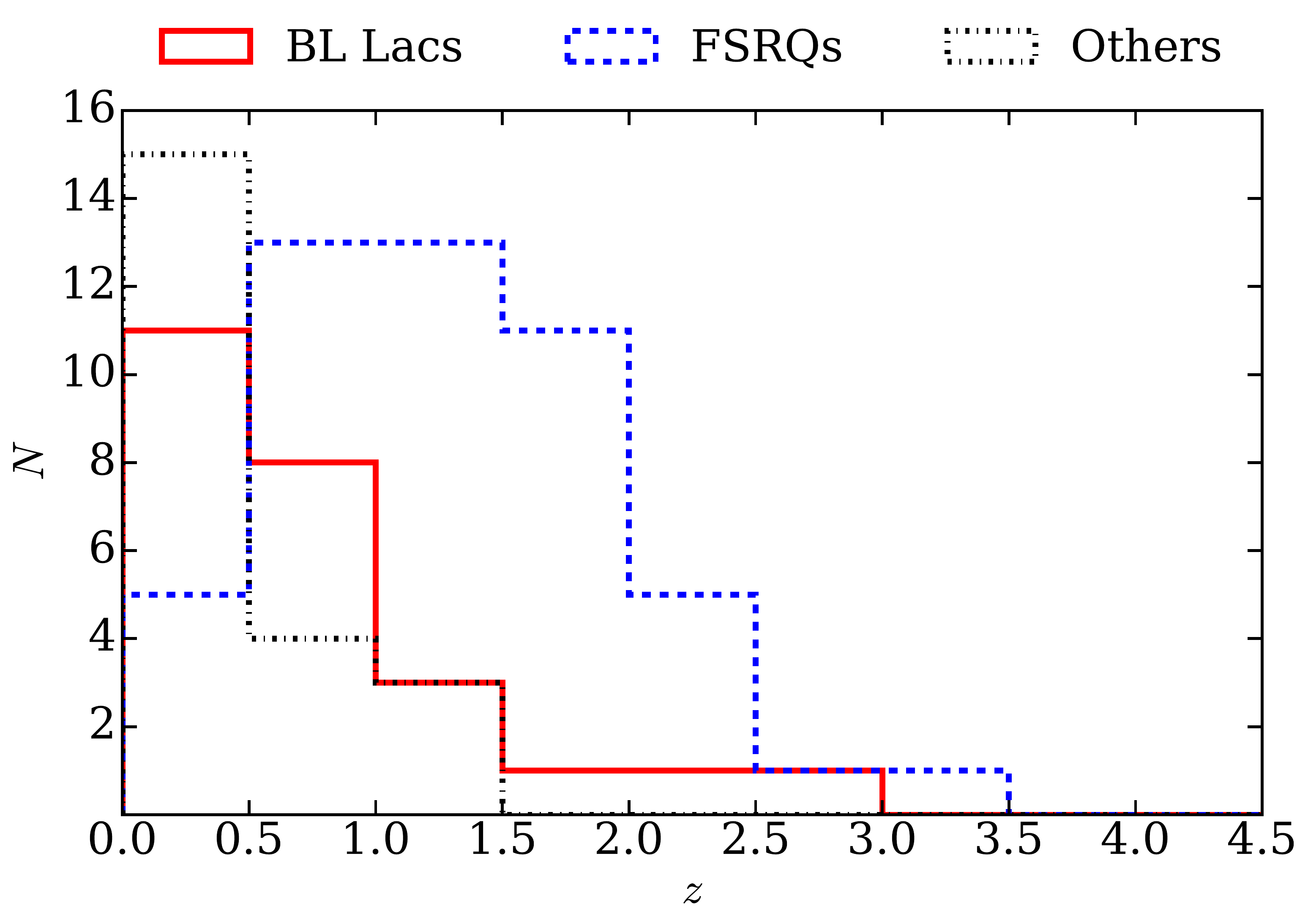}
        \caption{ }
        \label{fig:z-histogram}
    \end{subfigure}%
    \begin{subfigure}{0.33\textwidth}
                \includegraphics[width=\textwidth]{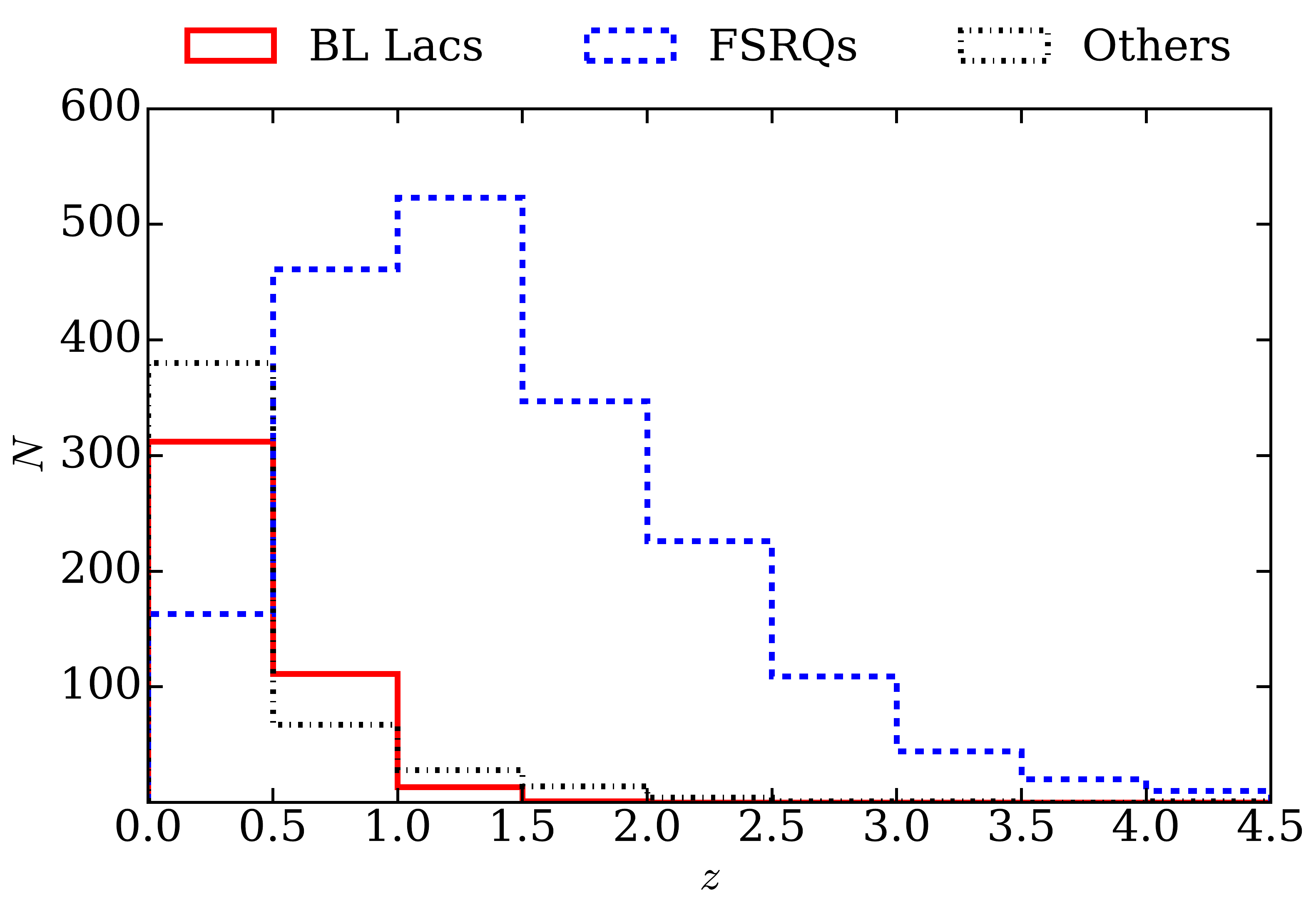}
        \caption{ }
        \label{fig:z-histogram-bzcat}
    \end{subfigure}%
    \begin{subfigure}{0.33\textwidth}
                \includegraphics[width=\textwidth]{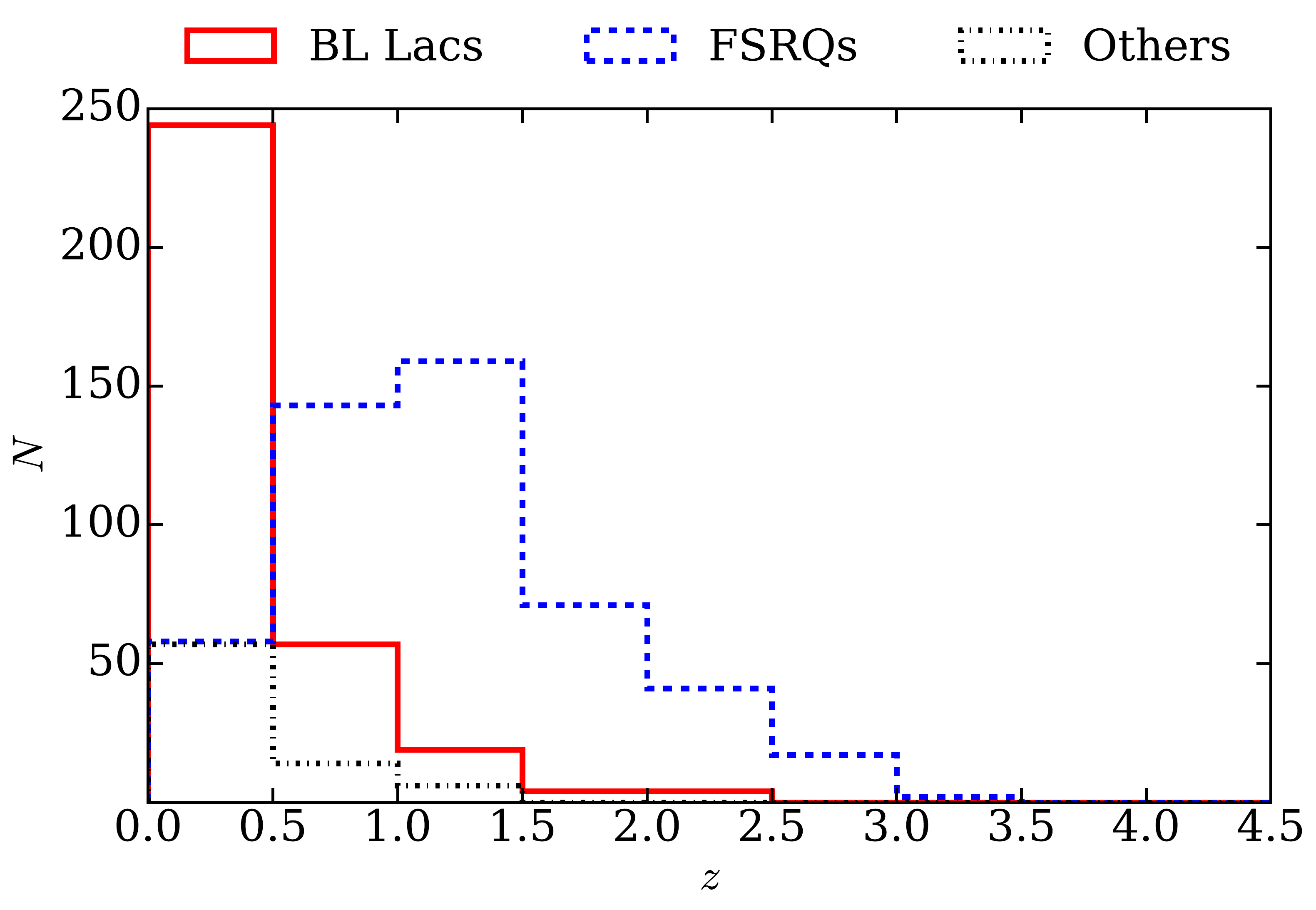}
        \caption{ }
        \label{fig:z-histogram-3lac}
    \end{subfigure}
    \caption{Distribution of the measured redshift values in (a) our sample compared to (b) BZCAT and (c) 3LAC is shown. Included in ``Others'' are, for example, blazars of uncertain type, BL Lac candidates, and galaxy-dominated BL Lacs.}
    \label{fig:z}
\end{figure*}

\subsection{Blazar variability} \label{sbs:biases}

Blazars can exhibit flux variability from radio to $\gamma$-ray energies \citep{2011ApJS..194...29R}, making it necessary to assess the impact of any inherent variability on the derived spectral indices and the strength of the radio-to-$\gamma$-ray correlation.

The surveys used to calculate the spectral index are non-simultaneous, so it is possible that the flux densities used to fit $\alpha$ change over time. However, there is less radio variability in blazars below the synchrotron peak than above it \citep{1998AdSpR..21...89U}. In support of this, \cite{bell2018} found that blazars do not seem to be significantly variable at low frequencies and \cite{1990MNRAS.246..123M} found little variability of 7C sources at \SI{151}{\mega\hertz}. \cite{2016sf2a.conf..373P} monitored PKS 2155-304, one of the brightest BL Lacs, while it was flaring and found only marginal variability at \SI{235}{\mega\hertz}. Furthermore, \cite{2015IAUS..313...95T} conducted a preliminary blazar monitoring programme with LOFAR at \SI{226}{\mega\hertz}, focussing on five blazars which exhibit strong gigahertz variability. The LOFAR light curves revealed a smooth behaviour (with some possible changes to the flux of the order of months). Hence, it is the NVSS flux densities which we expect to be most affected by variability, since this was the only catalogue we used $>\SI{325}{\mega\hertz}$. We included data from several megahertz surveys in the spectral modelling to minimise the influence of this possible variability, but the NVSS data are still the most influential when calculating $\alpha$.

The LDR1 and 3FGL catalogues are non-contemporaneous: LDR1 observations were made between $2014$ and $2015$ while 3FGL observations were integrated between the years $2008$ and $2012$. As a result, for any blazars which exhibit strong $\gamma$-ray variability, the data in 3FGL correspond to an average value and are more indicative of the non-flaring state. Since we do not expect the \SI{144}{\mega\hertz} or $\gamma$-ray data to be variable, we conclude that the non-simultaneity does not significantly impact any correlation between the radio and $\gamma$-ray bands.

\section{Results} \label{sec:results}

\subsection{Detection rate and redshift} \label{sbs:sampleproperties}

We identified LDR1 counterparts to $100\%$ of the high-energy sources (excluding UGS) and a summary of our results are given in Table~\ref{tab:loudness}. Information on the individual $98$ sources in our sample is presented in Table~\ref{long-table} at the end of this paper. In our sample, $48\%$ of sources are FSRQs, $25\%$ are BL Lacs, $8\%$ are blazars of uncertain type or BL Lac candidates, and $16\%$ are other source types (e.g. galaxy-dominated blazars, AGN, radio galaxies, and UGS).

Most (77/98) redshifts are the spectroscopic LDR1 values. The remainder are from BZCAT (6/98), the NASA/IPAC Extragalactic Database (6/98), 3LAC (1/98), or are LDR1 photometric estimates (6/98); two sources have no measured redshift. Obtaining photometric redshifts for blazars is challenging owing to the lack of reliable SED templates, but the LDR1 photometric redshifts are dominated by machine learning estimates which do not depend on such templates. The caption of Table \ref{long-table} contains a link to the CSV version of the table, which shows the origin of $z$ for each source.

Fig.~\ref{fig:z} shows the redshift distribution of our sample as well as the distributions of BZCAT and 3LAC. In BZCAT, $2\,842$ of the $3\,561$ sources have redshifts (see Fig.~\ref{fig:z-histogram-bzcat}), and in 3LAC, $896$ of the $1\,773$ sources have a measured redshift (see Fig.~\ref{fig:z-histogram-3lac}). The redshift distribution of our sample follows a similar trend to the BZCAT distribution, but in 3LAC there is a larger percentage of low-redshift BL Lacs. The FSRQ population is more distant than BL Lacs on average in all cases.

\subsection{Flux density and luminosity} \label{sbs:radioflux}

The \SI{144}{\mega\hertz} radio flux density, $S_{144\,\mathrm{MHz}}$, in our sample ranges from \SI{1.3}{\milli Jy} to \SI{14}{Jy}. The FSRQs have a higher median $S_{144\,\mathrm{MHz}}$ than the BL Lacs, as seen in Table~\ref{tab:loudness}. The median $S_{144\,\mathrm{MHz}}$ for $\gamma$-ray-detected sources (\SI{193 \pm 105}{mJy}) and for non-$\gamma$-ray-detected sources (\SI{203 \pm 19}{mJy}) are within error of each other.

\begin{figure}
        \includegraphics[width=\columnwidth]{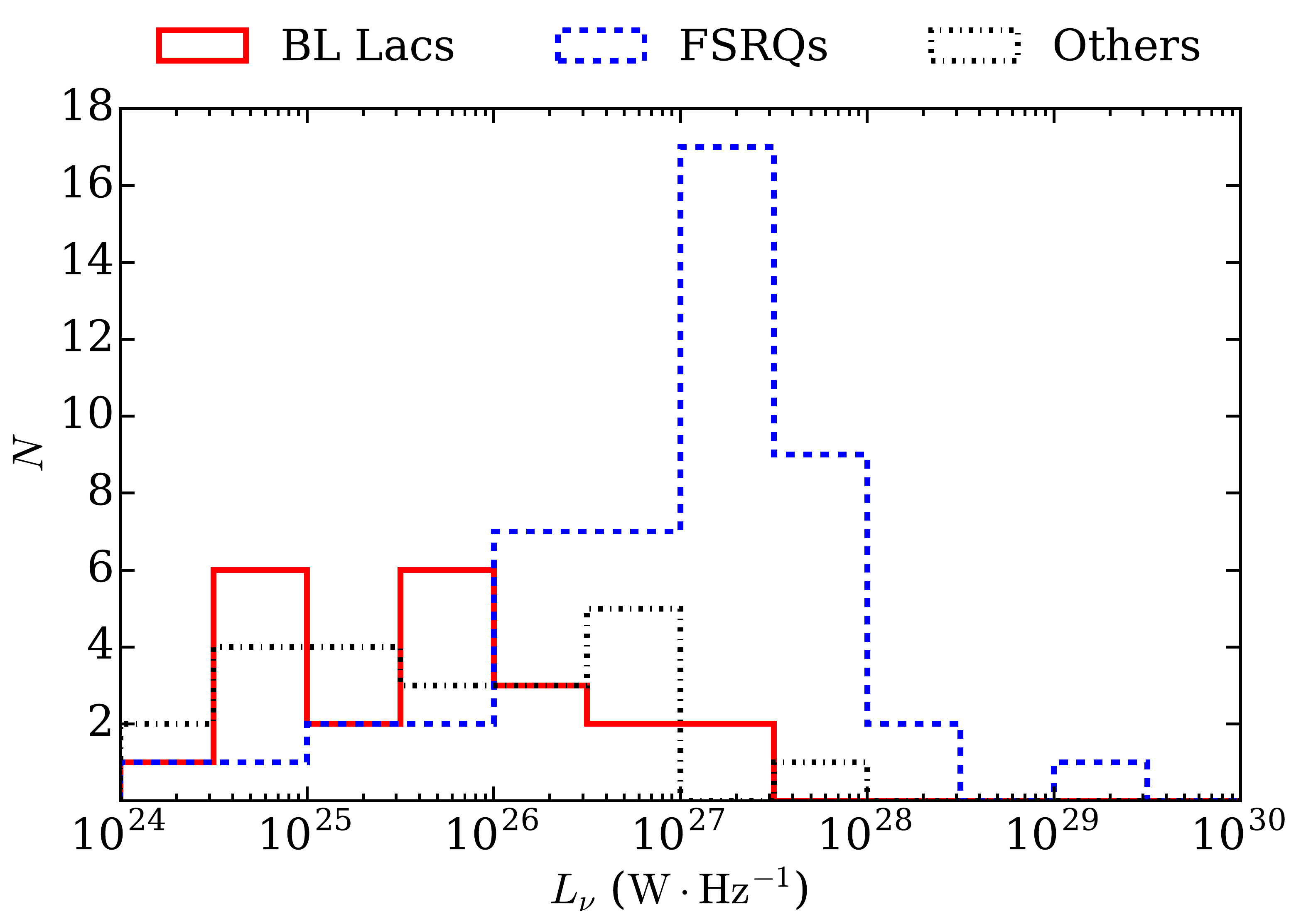}
    \caption{Radio luminosity distribution for our sample is shown.}
    \label{fig:radio_flux_histogram}
\end{figure}

We calculated the radio luminosity, $L_{\nu}$ (in \si{\watt\per\hertz}), according to
$$ L_{\nu }={\frac {S_{144\,\mathrm{MHz}}4\pi {d}^{2}}{(1+z)^{1+\alpha }}}, $$
where $S_{144\,\mathrm{MHz}}$ is in \si{\watt\per\metre\square\per\hertz} and $d$ is the luminosity distance in metres. Fig.~\ref{fig:radio_flux_histogram} shows the distribution of $L_{\nu}$ for our sample. The FSRQs span a broad range of $L_{\nu}$ while the BL Lacs are predominantly in the lower bins, as expected.

\subsection{Radio spectral index} \label{sbs:radiospectralindex}

\begin{figure}
        \includegraphics[width=\columnwidth]{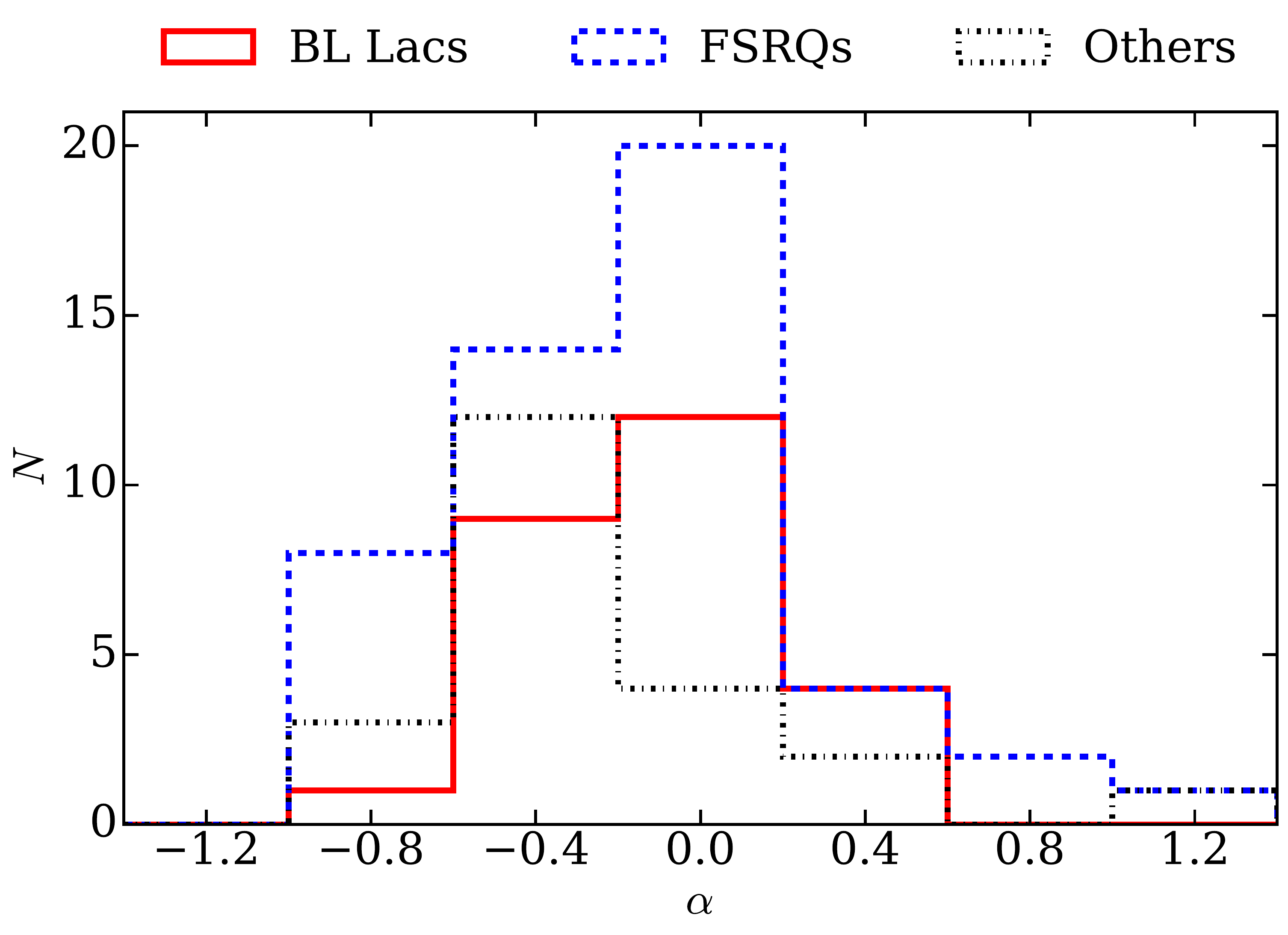}
    \caption{Distribution of the radio spectral indices for the sources in our sample is shown.}
    \label{fig:radio_alpha}
\end{figure}

Fig.~\ref{fig:radio_alpha} shows the radio spectral index distribution, and the average values are given in Table~\ref{tab:loudness}. The average $\alpha$ for our sample is $-0.17 \pm 0.14$, and this is much flatter than $\alpha$ for all sources in LDR1, which we expect to be $-0.8 \lesssim \alpha \lesssim -0.7$. Our results suggest both BL Lacs and FSRQs are flat even at megahertz frequencies. We found $\alpha = -0.11 \pm 0.17$ for the $\gamma$-ray sources, which is similar to the non-$\gamma$-ray-detected blazars where $\alpha = -0.21 \pm 0.16$. \cite{2016A&A...588A.141G} found $\alpha = -0.57 \pm 0.02$, which is steeper than the $\alpha$ we calculated. This can be explained by the fact that all blazars in the field were detected in this study, whereas \citeauthor{2016A&A...588A.141G} detected $36\%$ of blazars. This introduces a selection effect against flat or inverted-spectrum sources.

The contribution of the flat-spectrum core to the flux density is understood to decrease as the frequency decreases and, in the megahertz regime, the flux density is thought to be dominated by the extended emission in the radio lobes. However, the flatness of $\alpha$ suggests the beamed core emission is contributing somewhat to the low-frequency flux density. As our sample consists of blazars, Doppler boosting can lead to the core component appearing disproportionately brighter than the extended component.

\begin{figure}
        \includegraphics[width=\columnwidth]{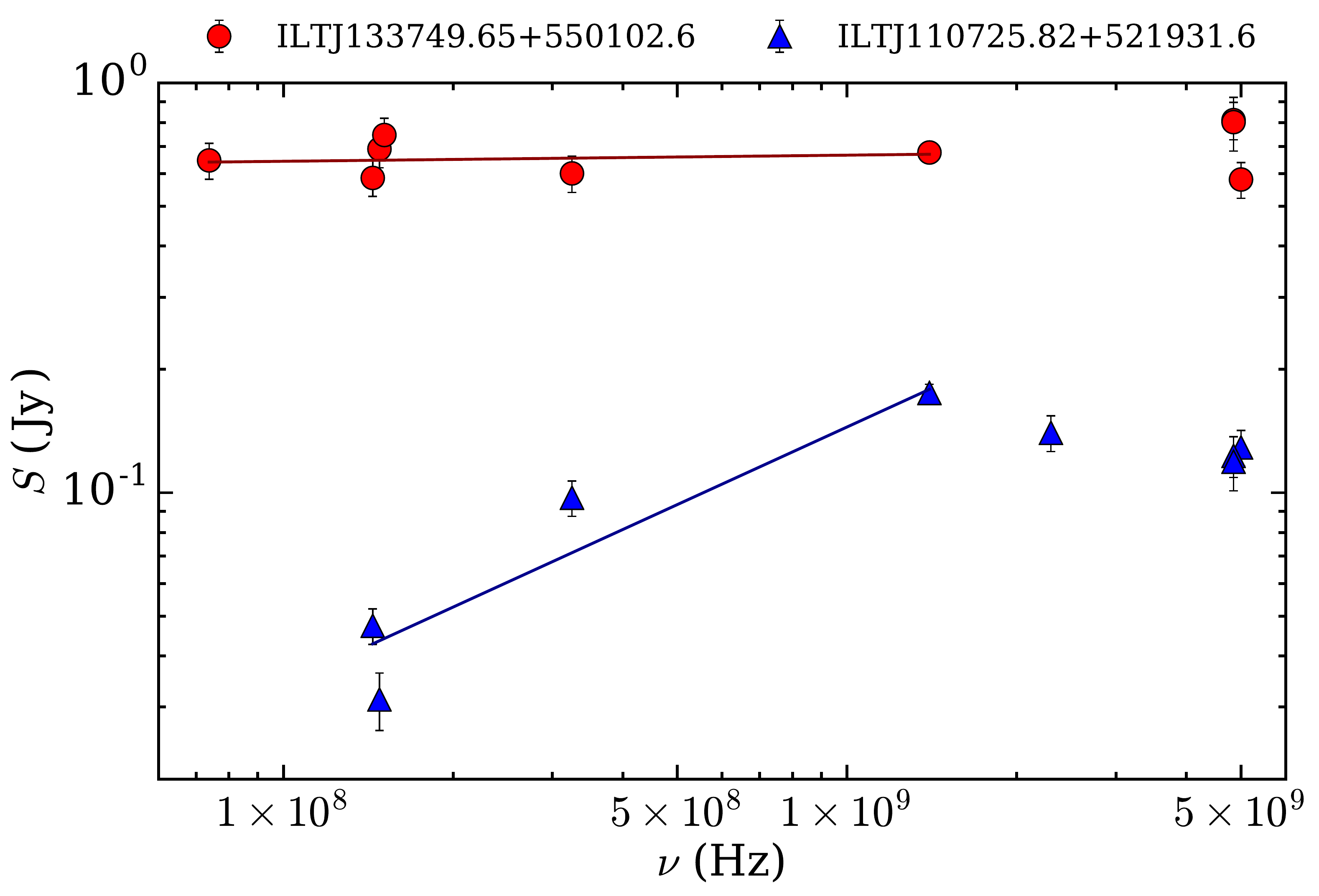}
    \caption{Flat spectrum source and peaked spectrum source from our sample. The spectral fits we derived using data up to \SI{1.4}{\giga\hertz} are shown as solid lines. Data from NED up to \SI{5}{\giga\hertz} have been plotted. The flatness of the radio spectrum for ILTJ133749.65+550102.6 is clear, as is the \si{\giga\hertz} peak for ILTJ110725.82+521931.6.}
    \label{fig:radio-spectra}
\end{figure}

The spectra for some sources in our sample appear to be gigahertz-peaked spectrum (GPS) sources \citep{2017ApJ...836..174C}. An example is seen in Fig.~\ref{fig:radio-spectra}, which shows the spectra from which $\alpha$ was derived for two sources in our sample. It has previously been argued that GPS quasars are flaring blazars \citep{2005A&A...432...31T} or intrinsically young radio sources \citep{1995A&A...302..317F}. % blazars in a dense gas environment \citep{2005JKAS...38..125B}

\subsection[Radio--gamma-ray connection]{Radio--$\gamma$-ray connection} \label{sbs:rgrc}

\begin{figure}
        \includegraphics[width=\columnwidth]{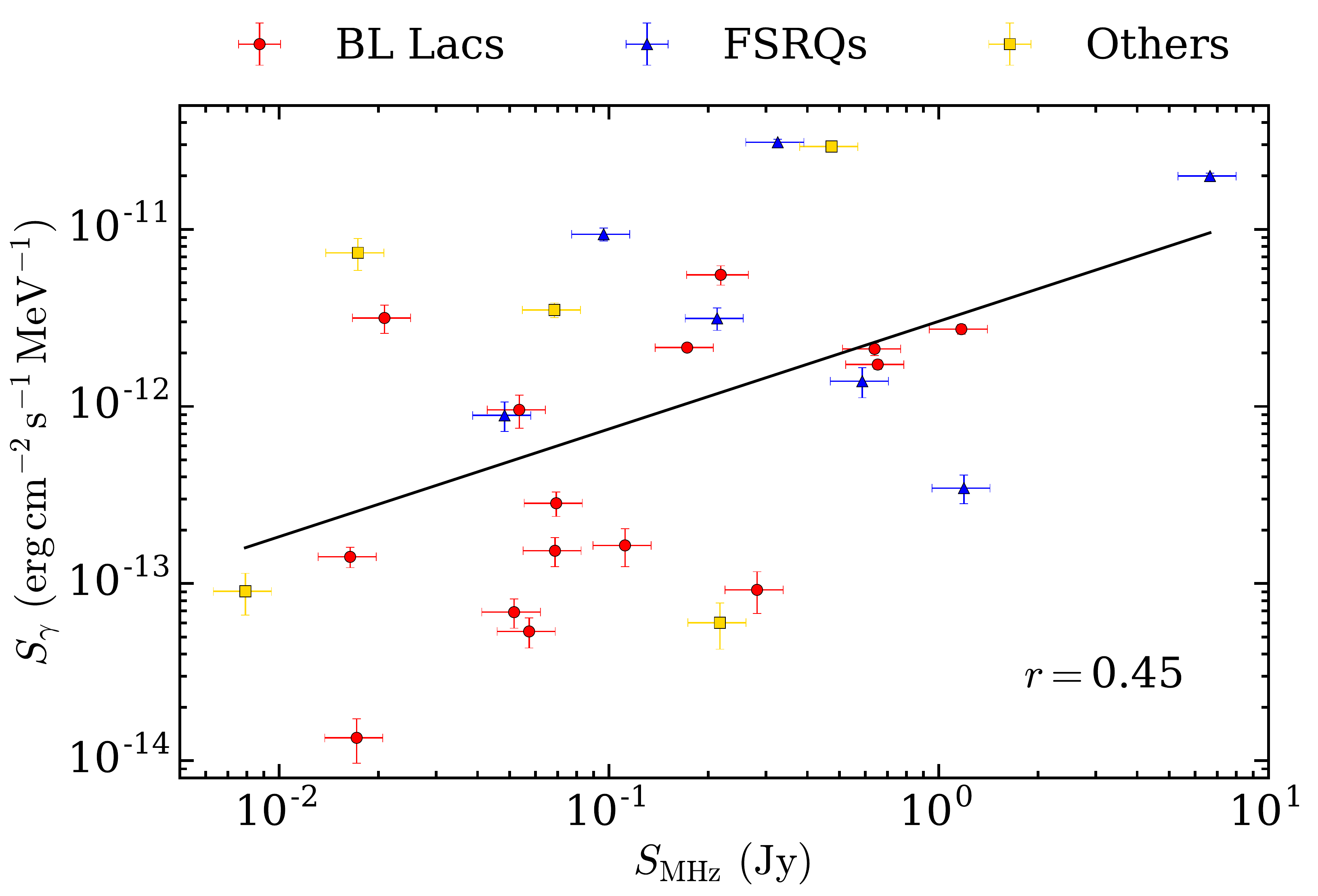}
    \caption{ $\gamma$-ray energy flux density is plotted against the radio flux density. A line was fit to the logarithms of the data. The radio flux density is at \SI{144}{\mega\hertz} and the $\gamma$-ray energy flux density measurements correspond to $\SI{100}{\mega\electronvolt}$.}
    \label{fig:log_gamma_vs_log_radio_3}
\end{figure}

Fig.~\ref{fig:log_gamma_vs_log_radio_3} shows the radio flux density plotted against the $\gamma$-ray energy flux for the $\gamma$-ray-detected sources. The $\gamma$-ray energy flux at $\SI{100}{MeV}$ was calculated from the integrated photon flux given in 3FGL using the $\gamma$-ray power-law spectral index. Two sources not in the 3LAC ``clean'' sample and 3C~303 were excluded. A linear fit to the logarithms of the flux yields a slope, and hence power-law index, of $m = 0.61 \pm 0.25$. We obtained a Pearson correlation coefficient, $r$, of $0.45$ with a null-hypothesis $p$-value of $0.019$. This marginally significant $p$-value is limited by our $N = 27$ sample size, and the sample sizes were too small to calculate the correlation with any meaningful significance for the BL Lac or FSRQ populations. This correlation also does not address the biases within the data.

\begin{figure*}
    \centering
    \begin{subfigure}{0.33\textwidth} % colour-diagrams-w1w2-w1-sample.pdf
                \includegraphics[width=\textwidth]{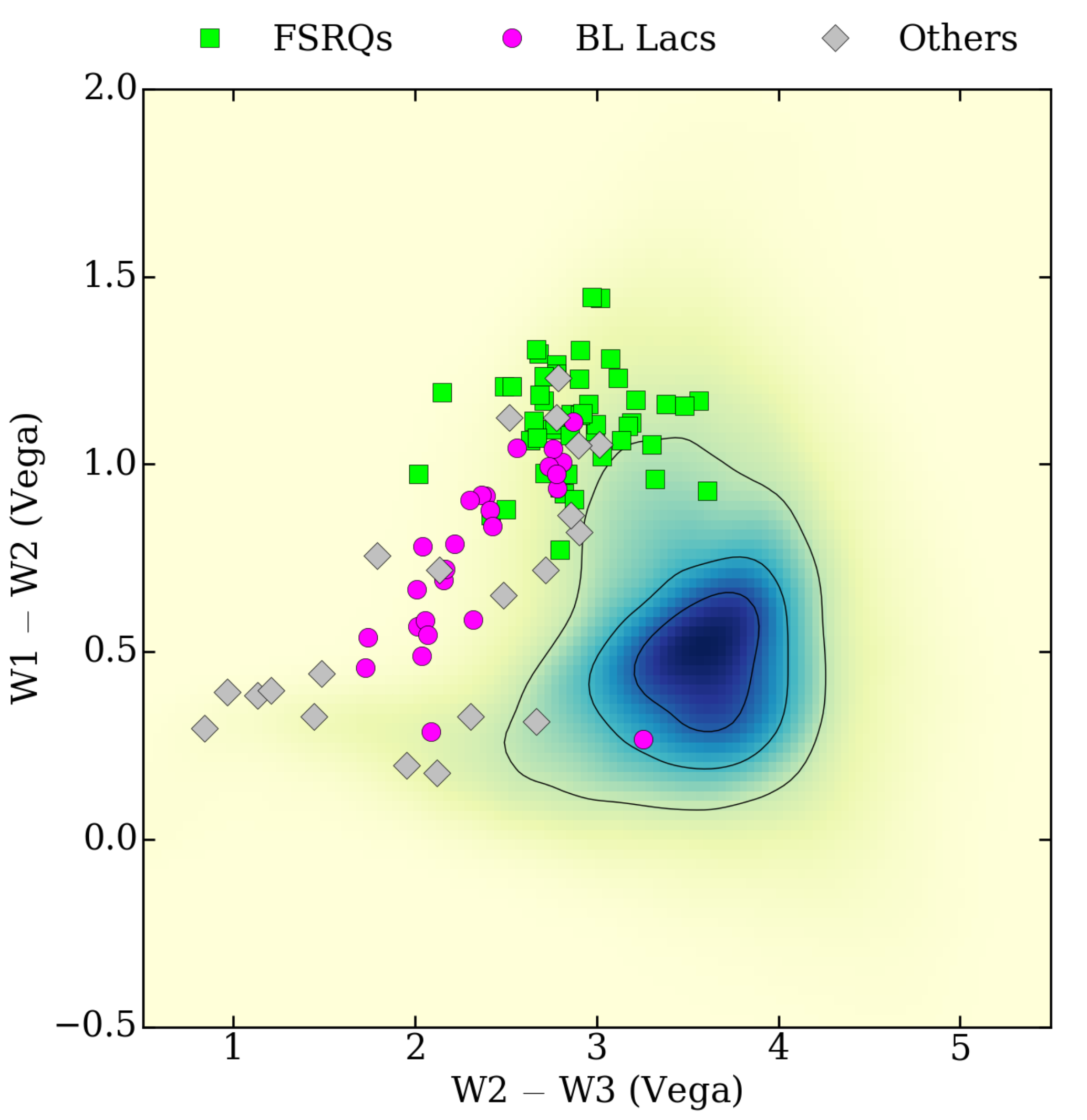}
        \caption{}
        \label{fig:colour1}
    \end{subfigure}%
    \begin{subfigure}{0.33\textwidth}
                \includegraphics[width=\textwidth]{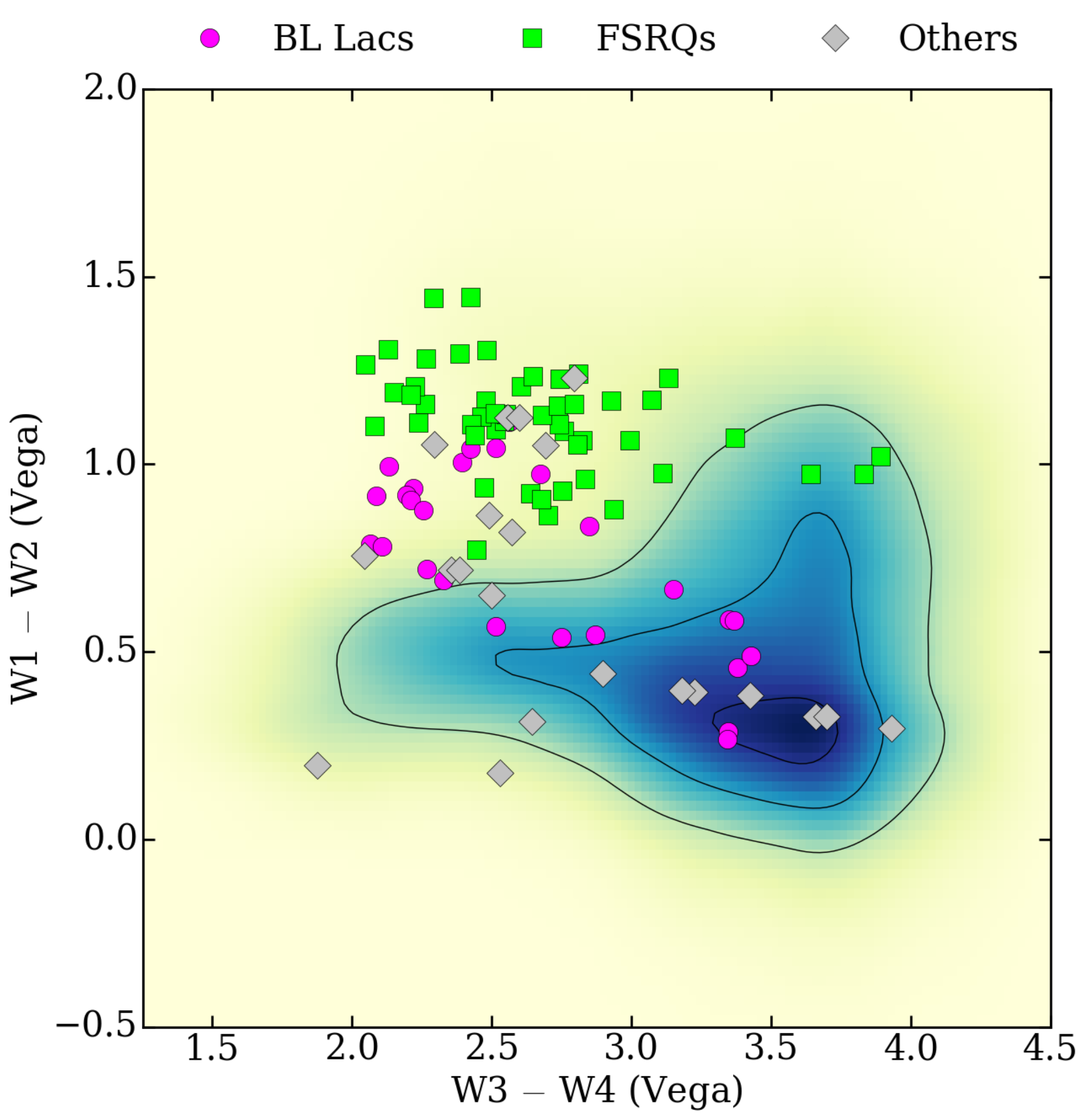}
        \caption{}
        \label{fig:colour2}
    \end{subfigure}%
    \begin{subfigure}{0.33\textwidth}
                \includegraphics[width=\textwidth]{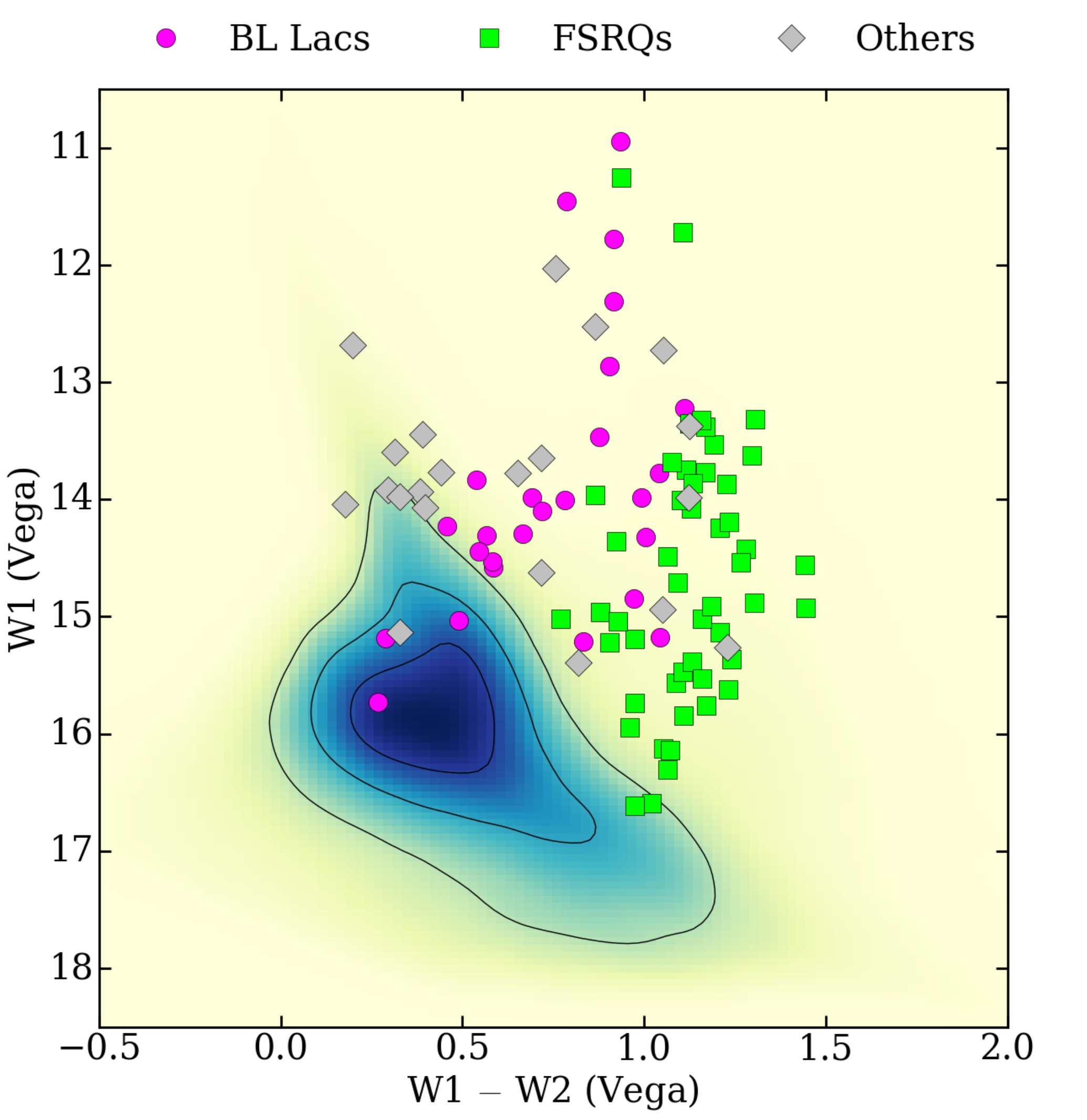}
        \caption{}
        \label{fig:colour3}
    \end{subfigure}
    \caption{Two-dimensional histograms show (a) the $W1 - W2$ vs. $W2 - W3$ and (b) the $W1 - W2$ vs. $W3 - W4$ colour-colour diagrams, and (c) the $W1 - W2$ vs. $W1 - W2$ colour-magnitude diagram. The IR colours for the entire LDR1 sample for which there is \textit{WISE} data available ($218\,595$ of $318\,520$ sources) is the two-dimensional histogram, with contours indicating the $25\%$, $50\%$, and $75\%$ levels. The points are the LDR1 sources from our sample for which there is \textit{WISE} data ($96$ of $98$ sources).}
    \label{fig:colour}
\end{figure*}

We then used the Monte Carlo correlation method outlined by \cite{2012ApJ...751..149P} in our radio-to-$\gamma$-ray analysis, which has also been used by \cite{2011ApJ...741...30A}. This method was designed for small samples affected by distance effects and subjective sample selection criteria. The data are randomised in luminosity space. \bred{This accounts for the fact that the radio and $\gamma$-ray flux densities appear to be correlated because of their common redshift. Then} the significance is measured in flux space to avoid Malmquist bias \citep{1997ApJ...476..572L}.

The \citeauthor{2012ApJ...751..149P} method gave the $r = 0.45 $ correlation a significance of $p = 0.069$. This is therefore suggestive of a correlation, although we cannot conclusively reject the null hypothesis that the radio and $\gamma$-ray luminosities of blazars are independent. Furthermore, this method provides a conservative estimate for small samples and so, while real correlations may not be verified, exaggerated significances are avoided in cases where there is insufficient evidence.

This correlation is weaker than the gigahertz radio-to-$\gamma$-ray connection for the same sample ($r = 0.57$, $p=0.002$), as we would expect, given that the emission is usually more diffuse at lower frequencies. % Nonetheless, the correlation establishes a link between the two components of the double-humped blazar SED, suggesting that the electrons producing the low-energy emission play a key role in producing the high-energy SED component. It is likely that these electrons which produce synchrotron radiation are involved in an inverse-Compton scattering process, although we cannot infer whether the seed photons originate internally or externally.

% The MHz emission is believed to be dominated by the older, more diffuse emission found in the radio lobes. However, the correlation implies that the beamed core emission that gives rise to the $\gamma$-rays is contributing to the low-frequency emission.

\subsection{Colour-colour diagrams}

In 2010, \textit{WISE} observed the sky at \SI{3.4}{\micro\metre} ($W1$), \SI{4.6}{\micro\metre} ($W2$), \SI{12}{\micro\metre} ($W3$), and \SI{22}{\micro\metre} ($W4$). These magnitudes are included in LDR1, from which we calculated the colours. Fig.~\ref{fig:colour} shows the $W1-W2-W3$ colour-colour and colour-magnitude diagrams. Sources in our sample are plotted over the LDR1 catalogue, where LDR1 sources are predominantly star-forming galaxies. It is clear that blazars populate distinct regions compared to LDR1 on each of these plots, but the blazar population is the most compact in Fig~\ref{fig:colour1}.

In Fig.~\ref{fig:colour1}, the \textit{WISE} blazar strip can be seen \citep{2011ApJ...740L..48M}. Generally, blazars are dominated by synchrotron emission in the infrared (IR) band. As a result, blazars have a distinct locus to that of the LDR1 sources, the majority of which of are dominated by a thermal component in the IR. The distribution of blazars in Fig.~\ref{fig:colour1} is in agreement with a power-law model for the IR spectrum. Moreover, BL Lacs and FSRQs also inhabit distinct regions on this colour-colour diagram, and the locations of these populations are consistent with the findings of \cite{2011ApJ...740L..48M}. Some blazars lie outside the blazar strip, and in this case, it is possible that there is a non-negligible thermal contribution to the IR emission from the host galaxy.

Fig.~\ref{fig:colour2} shows a different combination of IR colours, where the blazar population is also removed from the thermal LDR1 population. Fig.~\ref{fig:colour3} plots the $W1$ magnitude, which is the band with the highest sensitivity, against the $W1 - W2$ colour. The blazars appear brighter than LDR1 sources of the same colour as a result of Doppler boosting. The majority of blazars have $W1 - W2 \approx 1$, as noted by \cite{2012ApJ...748...68D}. This corresponds to an IR spectral index of $-1$ and suggests the synchrotron component peaks close to the \textit{WISE} measurements. Furthermore, \cite{2012ApJ...753...30S} used $W1 - W2 > 0.8$ as criterion to select for AGN, as this distinguishes between the AGN power-law spectra and the galactic black-body spectra. 

\begin{figure}
        \includegraphics[width=\columnwidth]{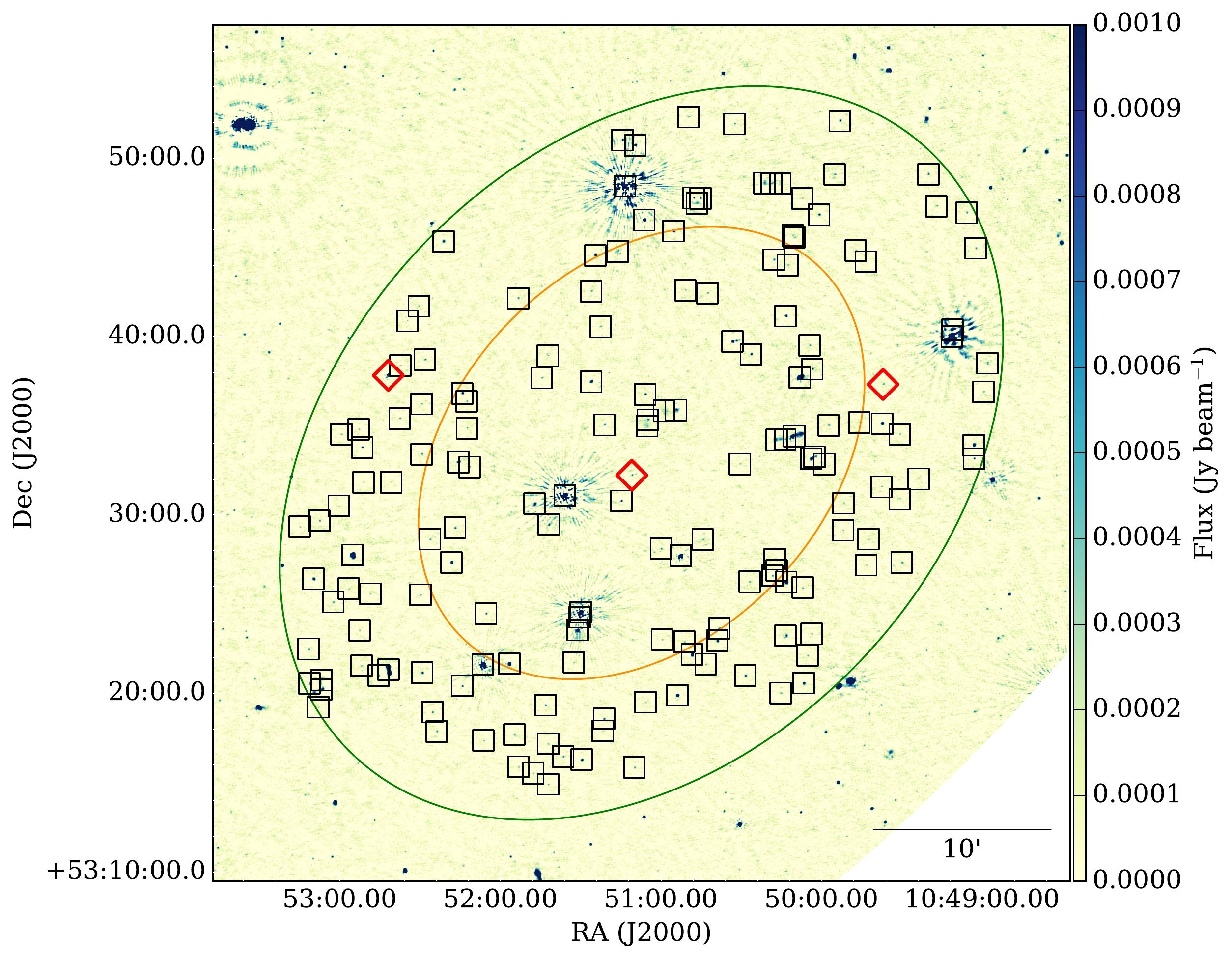}
    \caption{Unassociated $\gamma$-ray source 3FGL~J1051.0+5332 with $68\%$ (yellow) and $95\%$ (green) confidence ellipses is shown. Squares mark the LDR1 sources which lie within the $95\%$ confidence band; the three red diamonds are the sources which we assess to be the most likely match.}
    \label{fig:unassociated}
\end{figure}

The four 3FGL UGS within the LDR1 footprint are shown in Table~\ref{tab:ugs-1} alongside the number of LDR1 sources which fall within the 3FGL $95\%$ ellipse. This is illustrated for 3FGL~J1051.0+5332 in Fig.~\ref{fig:unassociated} where the semi-major and semi-minor axes are \SI{0.213}{\degree} and \SI{0.155}{\degree}, respectively. Colour information from \textit{WISE} has previously been used to classify UGS by \cite{2013ApJS..206...12D}. We also used the colour-colour diagrams to identify possible counterparts to the UGS on the basis that, statistically, these $\gamma$-ray sources are most likely to be blazars because blazars dominate the extragalactic $\gamma$-ray sky.  % \LEt{Please check for intended meaning.} %CHANGECHANGE: blazar type == blazars

\begin{table}
\centering
\caption{List of UGS within the LDR1 footprint, along with the number of LDR1 sources within the $95\%$ uncertainty ellipse. The 3FGL ellipse sizes vary considerably, leading to the large variation in the number of LDR1 matches. Also shown are the number of likely matches based on the colour information.}
\label{tab:ugs-1}
\begin{tabular*}{\columnwidth}{@{\extracolsep{\fill}}lcc@{}}
\toprule
\multicolumn{1}{c}{\textbf{UGS name}} & \textbf{\begin{tabular}[c]{@{}c@{}}LDR1 sources\\ within 95\%\end{tabular}} & \textbf{\begin{tabular}[c]{@{}c@{}}Likely matches\\ using colour data\end{tabular}} \\ \midrule
\;3FGL J1051.0+5332 & 166                                                        & 3                                                         \\
\;3FGL J1103.3+5239 & \,\;82                                                         & 1                                                         \\
\;3FGL J1231.6+4825 & \;\,29                                                         & 0                                                         \\
\;3FGL J1502.2+5553 & \;\,\;\,5                                                          & 0                                                         \\
\;\textbf{Total}    & \textbf{282}                                               & \textbf{4}                                                \\ \bottomrule
\end{tabular*}
\end{table}

\bred{We chose $96$ sources at random from LDR1 because colour information was available for $96$ sources in our sample.} We plotted these two populations on colour-colour diagrams \bred{(not shown)} and used an inverse-distance-weighted $k$-nearest neighbours ($k$-NN) algorithm (where $k = 3$) to identify which UGS matches were likely to be blazars. From the total $282$ possible LDR1 matches for the UGS, \bred{only four are likely to be blazars.} Three matched with one source, one UGS had just a single possibility, and two UGS had no possible matches remaining. The properties of these potential matches are given in Table~\ref{tab:ugs-2}.

% Of our sources, $96$ have colour information provided so we chose $96$ other sources at random from LDR1 which are not known blazars. We plotted these distributions on the colour-colour diagrams. We then used an inverse-distance-weighted $k$-nearest neighbours ($k$-NN) algorithm (where $k = 3$) to identify which UGS matches were likely to be blazars. From the total $272$ possible LDR1 matches, there were only four remaining. Three matched with one source, one UGS had just a single possibility, and two UGS had no possible matches remaining. The properties of these potential matches are given in Table~\ref{tab:ugs-2}.

Advantages of this $k$-NN method are that it assumes no prior knowledge of the region inhabited by blazars in colour space and that the implementation is straightforward. Identifying two of four possible counterparts is a promising return and based on experience the resulting matches seem plausible, but we will be able to better quantify the success of this method as the LoTSS footprint increases. We will also be able to test the reliability of this method against other supervised \citep[e.g. principal component analysis, as used by][]{2013ApJS..206...12D} and unsupervised (e.g. $k$-means clustering) machine learning techniques. However, these algorithms can only successfully identify quintessential blazars and those for which \textit{WISE} data are available. For the UGS in this study without a likely match, it is possible that the counterparts lie beyond the blazar strip, where the synchrotron radiation is not the dominant component at IR wavelengths.

\section{Conclusions} \label{sec:conclusions}

We examined the radio properties of the high-energy sources from BZCAT and 3FGL within LDR1. Because of their broadband nature, studying how blazars behave at low frequencies is essential to understanding how they operate. Historically, studying the low-frequency properties of blazars as a population has proven difficult because it has not been possible to identify low-frequency radio counterparts to these high-energy sources with the limited angular resolution and sensitivity of ${\sim}\SI{100}{\mega\hertz}$ surveys. The LDR1 catalogue addresses this technological gap and it is a marked improvement over even recent low-frequency surveys, such as MWACS and TGSS ADR1, in terms of angular resolution and sensitivity. As a result, we were able to find counterparts for all 3FGL and BZCAT sources in our field (excluding the UGS). % \LEt{Please check for intended meaning and fix as needed.}

Despite their poorly-constrained $\gamma$-ray position and the density of sources in the LDR1 field, we were able to identify possible radio counterparts for two of the four UGS within the LDR1 footprint using the \textit{WISE} colour information provided in the value-added catalogue. The radio spectral index was not available for most of the possible counterparts as the sources only had an LDR1 detection in the radio regime, but the availability of LoTSS in-band spectral indices in a future data release could help in matching these UGS.

\begin{table*}
\centering
\caption{Likely matches to the UGS are given. Only one source was detected in another survey (NVSS) which we used to calculate $\alpha$. There is no uncertainty listed as the fit is degenerate. We also list whether or not the LDR1 sources can be found in the Million Quasar Catalogue (MQC) \citep{2017yCat.7280....0F}, which is a compendium of quasars and high-confidence quasar candidates.}
\label{tab:ugs-2}
\begin{tabular*}{\linewidth}{@{\extracolsep{\fill}}cccccccc@{}}
\toprule
\textbf{3FGL name} & \textbf{LDR1 name}    & \textbf{Separation (\textdegree)} & \textbf{MQC} & \textit{\textbf{z}} & $\bm{\alpha}$ & \textbf{\textit{S}\textsubscript{MHz} (mJy)} \\ \midrule
\;3FGL J1051.0+5332  & ILTJ104931.19+533623.6 & $0.235$                                                                       & Yes          & $1.2582$              &          --     & \;\,$1.17\pm0.14$ \\
\;3FGL J1051.0+5332  & ILTJ105106.97+533143.0 & $0.023$                                                                       & Yes          & $1.1291$             &            --   & \;\,$0.85 \pm 0.19$ \\
\;3FGL J1051.0+5332  & ILTJ105238.00+533738.3 & $0.247$                                                                       & Yes          & $0.4025$              &      $-0.34$    & $15.23\pm 0.25$      \\
\;3FGL J1103.3+5239  & ILTJ110327.25+523425.1 & $0.09\;\,$                                                                        & No           & $0.3716$              &    --    & \;\,$1.06\pm 0.11$        \\ \bottomrule
\end{tabular*}
\end{table*}

The $100\%$ detection rate of blazars in this study, alongside the wealth of ancillary information in the value-added catalogue makes the LoTSS first data release an extremely useful resource in studying the low-frequency properties of these high-energy sources. Indeed, preliminary efforts suggest that it may possible to use LDR1 to discover new blazars in the field, for follow up with other instruments. We looked to use this $k$-NN method to identify sources in LDR1 which are possibly blazars. From the $218\,595$ sources with four \textit{WISE} colours, ${\sim}1\%$ fell within the blazar-populated space for all of the colour diagrams. This number could be cut down further by placing sensible limitations on the redshift and spectral index and this will be investigated in a future study.

In total, there are $1\,444$ 3FGL sources and $2\,138$ BZCAT sources in the northern hemisphere sky, which is the final goal of LoTSS in terms of sky coverage. As LoTSS progresses, we plan to revisit this work and evaluate the properties of blazar subclasses. The inherent variability of blazars will be an ever-present issue because the different observational methods for the radio and $\gamma$-ray regimes means that acquiring perfectly contemporaneous observations is challenging. But a larger sample size means we will be able to deduce general trends with more confidence and this should reduce the influence of flaring blazars. It is fortunate that LoTSS comes at a time when \textit{Fermi} is still operational because \textit{Fermi} is unrivaled with respect to $\gamma$-ray detections. Weaker $\gamma$-ray sources will be present in 4FGL, the next \textit{Fermi}-LAT catalogue which is due to be released in 2018. The radio counterparts of these sources will be fainter too, possibly with more high-frequency-peaked BL Lacs, and LoTSS will be a valuable resource when it comes to identifying these sources.

\begin{longtab}
\fontsize{7.2}{9.6}\selectfont
\begin{landscape}
\begin{longtable}[c]{@{}ccccccclcccc@{}}
\caption{Sources in our sample listed alongside some key parameters. An electronic version of this table is available at \url{https://github.com/mooneyse/LDR1-blazars}.}
\label{long-table}\\
\toprule
\textbf{Number}  &  \begin{tabular}[c]{@{}c@{}} \textbf{RA} \\ \textbf{(\textdegree)} \end{tabular}  &  \begin{tabular}[c]{@{}c@{}} \textbf{Dec.} \\ \textbf{(\textdegree)} \end{tabular}  & \begin{tabular}[c]{@{}c@{}} \textbf{LDR1} \\ \\  \end{tabular}            &  \begin{tabular}[c]{@{}c@{}} \textbf{BZCAT} \\ \\  \end{tabular}  &  \begin{tabular}[c]{@{}c@{}} \textbf{3FGL} \\ \\  \end{tabular}      &  \begin{tabular}[c]{@{}c@{}} \textbf{Classification} \\ \\  \end{tabular}  &  \begin{tabular}[c]{@{}c@{}} \;\;\;\,\textbf{\textit{z}} \\ \\  \end{tabular}  &  \begin{tabular}[c]{@{}c@{}} \textbf{\textit{S}\textsubscript{144\,MHz}}\;\;\;\; \\ \textbf{(mJy)}\;\;\;\; \end{tabular}  &  \begin{tabular}[c]{@{}c@{}} $\bm{\alpha}$ \\ \\  \end{tabular}  &  \begin{tabular}[c]{@{}c@{}}\textbf{\textit{S}\textsubscript{$\bm{\gamma}$} (0.1--100 GeV)} \\ \textbf{(10\textsuperscript{-12} ph cm\textsuperscript{-2} s MeV)} \end{tabular}  &  \begin{tabular}[c]{@{}c@{}} $\bm{\Gamma}$ \\ \\  \end{tabular}  \\* \midrule
\endfirsthead
\multicolumn{12}{c}%
{{\bfseries Table \thetable\ continued}} \\
\toprule
\textbf{Number}  &  \begin{tabular}[c]{@{}c@{}} \textbf{RA} \\ \textbf{(\textdegree)} \end{tabular}  &  \begin{tabular}[c]{@{}c@{}} \textbf{Dec.} \\ \textbf{(\textdegree)} \end{tabular}  & \begin{tabular}[c]{@{}c@{}} \textbf{LDR1} \\ \\  \end{tabular}            &  \begin{tabular}[c]{@{}c@{}} \textbf{BZCAT} \\ \\  \end{tabular}  &  \begin{tabular}[c]{@{}c@{}} \textbf{3FGL} \\ \\  \end{tabular}      &  \begin{tabular}[c]{@{}c@{}} \textbf{Classification} \\ \\  \end{tabular}  &  \begin{tabular}[c]{@{}c@{}} \;\;\;\,\textbf{\textit{z}} \\ \\  \end{tabular}  &  \begin{tabular}[c]{@{}c@{}} \textbf{\textit{S}\textsubscript{144\,MHz}}\;\;\;\; \\ \textbf{(mJy)}\;\;\;\; \end{tabular}  &  \begin{tabular}[c]{@{}c@{}} $\bm{\alpha}$ \\ \\  \end{tabular}  &  \begin{tabular}[c]{@{}c@{}}\textbf{\textit{S}\textsubscript{$\bm{\gamma}$} (0.1--100 GeV)} \\ \textbf{(10\textsuperscript{-12} ph cm\textsuperscript{-2} s MeV)} \end{tabular}  &  \begin{tabular}[c]{@{}c@{}} $\bm{\Gamma}$ \\ \\  \end{tabular}  \\* \midrule
\endhead
\bottomrule
\endfoot
\endlastfoot
$\;\,1$  &  $161.600121$     &  $53.907374$     &  ILTJ104624.03+535426.5  &  5BZQJ1046+5354  &  ---                &  FSRQ                     & 1.712 &  $130 \pm 26$  &  $-0.03 \pm 0.17$  &  ---                    &   ---               \\
$\;\,2$  &  $161.940807$     &  $54.627993$     &  ILTJ104745.79+543740.7  &  5BZBJ1047+5437  &  ---                &  BL Lac Candidate         & 0.622 & \;\, $4 \pm 1$  & \;\, $0.21 \pm 0.2$  &  ---                    &   ---               \\
$\;\,3$  &  $162.816848$     &  $46.738246$     &  ILTJ105116.04+464417.6  &  5BZUJ1051+4644  &  ---                &  Uncertain                & 1.419 & \;\, $718 \pm 152$  &  $-0.41 \pm 0.13$  &  ---                    &   ---               \\
$\;\,4$  &  $163.488478$     &  $54.031036$     &  ILTJ105357.23+540151.7  &  5BZQJ1053+5401  &  ---                &  FSRQ                     & 0.998 & \;\, $992 \pm 198$  &  $-0.61 \pm 0.13$  &  ---                    &   ---               \\
$\;\,5$  &  $164.656869$     &  $56.469764$     &  ILTJ105837.65+562811.1  &  5BZBJ1058+5628  &  3FGL J1058.6+5627  &  BL Lac                   & 0.143 &  $173 \pm 35$  & \;\, $0.09 \pm 0.13$  & \num{2.27(8)e-12} &  $1.945 \pm 0.025$  \\
$\;\,6$  &  $166.279471$     &  $46.888549$     &  ILTJ110507.07+465318.7  &  5BZGJ1105+4653  &  ---                &  Galaxy-dominated BL Lac  & 0.112 & \;\, $73 \pm 15$  &  $-0.19 \pm 0.2$ \;\, &  ---                    &   ---               \\
$\;\,7$  &  $166.76993$\,\;  &  $50.17706$\,\;  &  ILTJ110704.78+501037.4  &  5BZBJ1107+5010  &  ---                &  BL Lac                   & 0.706 & \;\, $35 \pm 7$ \;\, & \;\, $0.1 \pm 0.26$  &  ---                    &   ---               \\
$\;\,8$  &  $166.857565$     &  $52.325458$     &  ILTJ110725.82+521931.6  &  5BZQJ1107+5219  &  ---                &  FSRQ                     & 0.945 & \;\, $47 \pm 9$ \;\, & \;\, $0.64 \pm 0.21$  &  ---                    &   ---               \\
$\;\,9$  &  $167.773752$     &  $52.463669$     &  ILTJ111105.70+522749.2  &  5BZQJ1111+5227  &  ---                &  FSRQ                     & 1.284 &  $456 \pm 91$  &  $-0.63 \pm 0.12$  &  ---                    &   ---               \\
$10$     &  $169.488205$     &  $53.931894$     &  ILTJ111757.17+535554.8  &  5BZBJ1117+5355  &  3FGL J1117.9+5355  &  BL Lac                   & 2.953 &  $16 \pm 3$  &  $-0.13 \pm 0.11$  & \num{1.52(16)e-13} &   $1.930 \pm 0.076$ \\
$11$     &  $170.095916$     &  $54.074222$     &  ILTJ112023.02+540427.2  &  5BZQJ1120+5404  &  ---                &  FSRQ                     & 0.923 &  $1025 \pm 205$  &  $-0.61 \pm 0.12$  &  ---                    &   ---               \\
$12$     &  $171.056412$     &  $51.563419$     &  ILTJ112413.54+513348.3  &  5BZGJ1124+5133  &  ---                &  Galaxy-dominated BL Lac  & 0.235 & \;\, $87 \pm 17$  &  $-0.27 \pm 0.22$  &  ---                    &   ---               \\
$13$     &  $171.224277$     &  $49.56941$\,\;  &  ILTJ112453.83+493409.8  &  5BZBJ1124+4934  &  3FGL J1124.9+4932  &  BL Lac                   & 0.908 &  $17 \pm 3$  & \;\, $0.04 \pm 0.12$  & \num{1.68(34)e-14} &   $1.803 \pm 0.154$ \\
$14$     &  $174.509231$     &  $48.981876$     &  ILTJ113802.22+485854.7  &  ---             &  3FGL J1138.2+4905  &  Uncertain                & 1.305 &  $17 \pm 3$  & \;\, $0.38 \pm 0.11$  & \num{3.98(71)e-12} &   $2.851 \pm 0.182$ \\
$15$     &  $174.589363$     &  $47.755044$     &  ILTJ113821.45+474518.1  &  5BZQJ1138+4745  &  ---                &  FSRQ                     & 0.77 & \;\, $968 \pm 194$  &  $-0.52 \pm 0.13$  &  ---                    &   ---               \\
$16$     &  $175.199387$     &  $46.366832$     &  ILTJ114047.85+462200.6  &  5BZQJ1140+4622  &  ---                &  FSRQ                     & 0.115 &  $182 \pm 36$  &  $-0.33 \pm 0.17$  &  ---                    &   ---               \\
$17$     &  $175.988012$     &  $51.041951$     &  ILTJ114357.12+510231.0  &  5BZBJ1143+5102  &  ---                &  BL Lac                   &  \;\;---  & \;\, $91 \pm 18$  &  $-0.43 \pm 0.26$  &  ---                    &   ---               \\
$18$     &  $176.684008$     &  $53.945298$     &  ILTJ114644.16+535643.0  &  5BZQJ1146+5356  &  ---                &  FSRQ                     & 2.209 &  $411 \pm 82$  & \;\, $0.02 \pm 0.15$  &  ---                    &   ---               \\
$19$     &  $177.235842$     &  $52.906973$     &  ILTJ114856.60+525425.1  &  5BZQJ1148+5254  &  ---                &  FSRQ                     & 1.632 &  $221 \pm 44$  &  $-0.38 \pm 0.13$  &  ---                    &   ---               \\
$20$     &  $177.500649$     &  $55.47248$\,\;  &  ILTJ115000.16+552820.9  &  5BZGJ1150+5528  &  ---                &  Galaxy-dominated BL Lac  & 0.138 &  $156 \pm 31$  & \;\, $0.01 \pm 0.15$  &  ---                    &   ---               \\
$21$     &  $178.139215$     &  $49.660091$     &  ILTJ115233.41+493936.3  &  5BZQJ1152+4939  &  ---                &  FSRQ                     & 1.093 & \;\, $626 \pm 125$  &  $-0.8 \pm 0.14$  &  ---                    &   ---               \\
$22$     &  $178.351974$     &  $49.519345$     &  ILTJ115324.47+493109.6  &  5BZQJ1153+4931  &  3FGL J1153.4+4932  &  FSRQ                     & 0.334 & \;\, $6651 \pm 1331$  &  $-0.68 \pm 0.13$  & \num{1.45(5)e-11} &   $2.379 \pm 0.030$ \\
$23$     &  $179.611407$     &  $48.421098$     &  ILTJ115826.74+482515.9  &  5BZQJ1158+4825  &  ---                &  FSRQ                     & 2.036 &  $399 \pm 80$  &  $-0.21 \pm 0.17$  &  ---                    &   ---               \\
$24$     &  $180.191246$     &  $47.973316$     &  ILTJ120045.90+475823.9  &  5BZGJ1200+4758  &  ---                &  Galaxy-dominated BL Lac  & 0.27 &  $290 \pm 58$  &  $-0.88 \pm 0.13$  &  ---                    &   ---               \\
$25$     &  $180.877599$     &  $48.052322$     &  ILTJ120330.62+480308.3  &  5BZQJ1203+4803  &  ---                &  FSRQ                     & 0.343 & \;\, $77 \pm 15$  &  $-0.03 \pm 0.21$  &  ---                    &   ---               \\
$26$     &  $180.881573$     &  $54.502488$     &  ILTJ120331.58+543008.9  &  5BZBJ1203+5430  &  ---                &  BL Lac                   & 0.478 &  $172 \pm 34$  &  $-0.34 \pm 0.13$  &  ---                    &   ---               \\
$27$     &  $182.226107$     &  $54.699613$     &  ILTJ120854.27+544158.6  &  5BZQJ1208+5441  &  3FGL J1208.7+5442  &  FSRQ                     & 1.343 &  $325 \pm 65$  & \;\, $0.15 \pm 0.21$  & \num{2.01(7)e-11} &   $2.543 \pm 0.035$ \\
$28$     &  $183.253203$     &  $51.493158$     &  ILTJ121300.77+512935.3  &  5BZBJ1213+5129  &  3FGL J1212.6+5135  &  BL Lac                   & 0.796 &  $112 \pm 22$  &  $-0.49 \pm 0.17$  & \num{1.38(29)e-13} &   $2.191 \pm 0.150$ \\
$29$     &  $183.752901$     &  $50.037853$     &  ILTJ121500.70+500216.2  &  5BZBJ1215+5002  &  3FGL J1215.0+5002  &  BL Lac                   & 1.545 & \;\, $57 \pm 11$  & \;\, $0.22 \pm 0.15$  & \num{5.84(88)e-14} &   $1.919 \pm 0.111$ \\
$30$     &  $183.79149$\,\;  &  $46.454279$     &  ILTJ121509.96+462715.4  &  5BZQJ1215+4627  &  ---                &  FSRQ                     & 0.72 & \;\, $987 \pm 198$  &  $-0.57 \pm 0.12$  &  ---                    &   ---               \\
$31$     &  $184.402825$     &  $51.919516$     &  ILTJ121736.68+515510.2  &  5BZQJ1217+5155  &  ---                &  FSRQ                     & 2.225 & \;\, $79 \pm 16$  &  $-0.01 \pm 0.18$  &  ---                    &   ---               \\
$32$     &  $184.776916$     &  $48.499021$     &  ILTJ121906.46+482956.4  &  5BZQJ1219+4829  &  ---                &  FSRQ                     & 1.071 & \;\, $678 \pm 136$  &  $-0.02 \pm 0.16$  &  ---                    &   ---               \\
$33$     &  $185.282548$     &  $47.708018$     &  ILTJ122107.81+474228.8  &  5BZGJ1221+4742  &  ---                &  Galaxy-dominated BL Lac  & 0.21 &  $114 \pm 23$  &  $-0.5 \pm 0.19$  &  ---                    &   ---               \\
$34$     &  $185.912332$     &  $46.188986$     &  ILTJ122338.96+461120.3  &  5BZQJ1223+4611  &  ---                &  FSRQ                     & 1.012 & \;\, $516 \pm 103$  &  $-0.22 \pm 0.16$  &  ---                    &   ---               \\
$35$     &  $185.970166$     &  $46.846568$     &  ILTJ122352.84+465047.6  &  5BZGJ1223+4650  &  ---                &  Galaxy-dominated BL Lac  & 0.261 &  $12 \pm 2$  & \;\, $0.03 \pm 0.2$  &  ---                    &   ---               \\
$36$     &  $186.04125$\,\;  &  $50.032069$     &  ILTJ122409.90+500155.4  &  5BZQJ1224+5001  &  3FGL J1224.5+4957  &  FSRQ                     & 1.065 & \;\, $48 \pm 10$  &  $-0.05 \pm 0.13$  & \num{6.08(101)e-13} &   $2.463 \pm 0.131$ \\
$37$     &  $186.277633$     &  $48.576511$     &  ILTJ122506.63+483435.4  &  5BZUJ1225+4834  &  ---                &  Uncertain                & 0.647 &  $393 \pm 79$  &  $-0.76 \pm 0.13$  &  ---                    &   ---               \\
$38$     &  $186.982246$     &  $49.549112$     &  ILTJ122755.74+493256.8  &  5BZQJ1227+4932  &  ---                &  FSRQ                     & 0.551 &  $256 \pm 51$  &  $-0.12 \pm 0.17$  &  ---                    &   ---               \\
$39$     &  $187.215764$     &  $48.967037$     &  ILTJ122851.78+485801.3  &  5BZQJ1228+4858  &  3FGL J1228.7+4857  &  FSRQ                     & 1.722 &  $1194 \pm 239$  &  $-0.55 \pm 0.13$  & \num{2.53(41)e-13} &   $2.365 \pm 0.124$ \\
$40$     &  $188.144944$     &  $48.359178$     &  ILTJ123234.79+482133.0  &  5BZQJ1232+4821  &  ---                &  FSRQ                     & 1.588 & \;\, $656 \pm 136$  &  $-0.38 \pm 0.13$  &  ---                    &   ---               \\
$41$     &  $188.454997$     &  $50.439943$     &  ILTJ123349.20+502623.8  &  5BZGJ1233+5026  &  ---                &  Galaxy-dominated BL Lac  & 0.207 & \;\, $997 \pm 199$  &  $-0.58 \pm 0.11$  &  ---                    &   ---               \\
$42$     &  $188.555614$     &  $47.897647$     &  ILTJ123413.35+475351.5  &  5BZQJ1234+4753  &  ---                &  FSRQ                     & 0.374 &  $255 \pm 51$  & \;\, $0.13 \pm 0.15$  &  ---                    &   ---               \\
$43$     &  $188.878317$     &  $52.474719$     &  ILTJ123530.80+522828.9  &  5BZQJ1235+5228  &  ---                &  FSRQ                     & 1.653 & \;\, $61 \pm 12$  & \;\, $0.25 \pm 0.18$  &  ---                    &   ---               \\
$44$     &  $189.532619$     &  $53.431541$     &  ILTJ123807.83+532553.5  &  5BZUJ1238+5325  &  ---                &  Uncertain                & 0.347 & \;\, $578 \pm 116$  &  $-0.51 \pm 0.21$  &  ---                    &   ---               \\
$45$     &  $189.629331$     &  $54.114188$     &  ILTJ123831.04+540651.0  &  5BZGJ1238+5406  &  ---                &  Galaxy-dominated BL Lac  & 0.224 &  $112 \pm 22$  &  $-0.4 \pm 0.15$  &  ---                    &   ---               \\
$46$     &  $190.784689$     &  $52.213247$     &  ILTJ124308.33+521247.6  &  5BZGJ1243+5212  &  ---                &  Galaxy-dominated BL Lac  & 0.2 &  $148 \pm 30$  &  $-0.24 \pm 0.29$  &  ---                    &   ---               \\
$47$     &  $192.142403$     &  $51.468464$     &  ILTJ124834.18+512806.4  &  5BZBJ1248+5128  &  3FGL J1248.0+5130  &  BL Lac                   & 0.351 &  $282 \pm 56$  &  $-0.41 \pm 0.12$  & \num{7.92(175)e-14} &   $2.163 \pm 0.170$ \\
$48$     &  $193.30064$\,\;  &  $53.020266$     &  ILTJ125312.15+530112.9  &  5BZBJ1253+5301  &  3FGL J1253.2+5300  &  BL Lac                   & 0.178 &  $1172 \pm 235$  &  $-0.45 \pm 0.14$  & \num{2.74(12)e-12} &   $1.995 \pm 0.046$ \\
$49$     &  $194.161109$     &  $53.573331$     &  ILTJ125638.67+533423.9  &  ---             &  3FGL J1256.7+5328  &  Uncertain                & 0.591 & \;\, $530 \pm 106$  &  $-0.48 \pm 0.13$  & \num{1.43(25)e-12} &   $2.639 \pm 0.125$ \\
$50$     &  $194.604893$     &   $51.706896$    &  ILTJ125825.17+514224.8  &  ---             &  3FGL J1258.7+5137  &   Uncertain               & 0.441 &  $217 \pm 43$  &  $-0.39 \pm 0.26$  & \num{5.19(125)e-14} &   $2.159 \pm 0.189$ \\
$51$     &  $195.135991$     &   $53.853585$    &  ILTJ130032.64+535112.9  &  5BZBJ1300+5351  &   ---               &   BL Lac                  & 0.644 &  $95 \pm 19$  &  $-0.47 \pm 0.16$  &  ---                    &   ---               \\
$52$     &  $195.171868$     &   $50.493486$    &  ILTJ130041.25+502936.5  &  5BZQJ1300+5029  &   ---               &   FSRQ                    & 1.564 &  $839 \pm 168$  &  $-0.33 \pm 0.13$  &  ---                    &   ---               \\
$53$     &  $195.571269$     &   $48.321771$    &  ILTJ130217.10+481918.3  &  5BZQJ1302+4819  &   ---               &   FSRQ                    & 0.874 &  $171 \pm 34$  &  $-0.09 \pm 0.22$  &  ---                    &   ---               \\
$54$     &  $195.730643$     &   $50.938323$    &  ILTJ130255.35+505617.9  &  5BZBJ1302+5056  &   ---               &   BL Lac                  & 0.686 &  $3 \pm 1$  &  $-0.07 \pm 0.2$  &  ---                    &   ---               \\
$55$     &  $196.513726$     &   $55.495609$    &  ILTJ130603.29+552944.1  &  5BZQJ1306+5529  &   ---               &   FSRQ                    & 1.6 &  $142 \pm 28$  &  $-0.07 \pm 0.13$  &  ---                    &   ---               \\
$56$     &   $196.624684$    &   $47.692345$    &  ILTJ130629.92+474132.4  &  5BZQJ1306+4741  &   ---               &   FSRQ                    & 2.543 &  $52 \pm 10$  &  $-0.01 \pm 0.15$  &  ---                    &   ---               \\
$57$     &   $197.213143$    &   $47.498457$    &  ILTJ130851.15+472954.4  &  5BZQJ1308+4729  &   ---               &   FSRQ                    & 0.884 &  $332 \pm 66$  &  $-0.65 \pm 0.13$  &  ---                    &   ---               \\
$58$     &   $197.289151$    &   $55.959588$    &  ILTJ130909.40+555734.5  &  5BZQJ1309+5557  &   ---               &   FSRQ                    & 1.631 &  $292 \pm 58$  &  $-0.03 \pm 0.21$  &  ---                    &   ---               \\
$59$     &   $197.722941$    &   $46.897592$    &  ILTJ131053.51+465351.3  &  5BZQJ1310+4653  &   ---               &   FSRQ                    & 1.043 &  $74 \pm 15$  & \;\, $0.23 \pm 0.11$  &  ---                    &   ---               \\
$60$     &   $197.762372$    &   $55.231595$    &  ILTJ131102.97+551353.7  &  5BZQJ1311+5513  &  3FGL J1310.7+5515  &   FSRQ                    & 0.926 &  $587 \pm 118$  &  $-0.18 \pm 0.17$  & \num{9.09(155)e-13} &   $2.527 \pm 0.138$ \\
$61$     &   $198.043184$    &   $48.1564$      &  ILTJ131210.36+480923.0  &  5BZQJ1312+4809  &   ---               &   FSRQ                    & 0.715 &  $552 \pm 110$  &  $-0.49 \pm 0.17$  &  ---                    &   ---               \\
$62$     &   $198.180562$    &   $48.475364$    &  ILTJ131243.33+482831.3  &  5BZUJ1312+4828  &  3FGL J1312.7+4828  &   Uncertain               & 0.489 &  $474 \pm 95$  &  $-0.28 \pm 0.15$  & \num{2.59(7)e-11} &   $2.131 \pm 0.032$ \\
$63$     &   $199.876209$    &   $48.850909$    &  ILTJ131930.29+485103.2  &  5BZQJ1319+4851  &   ---               &   FSRQ                    & 1.168 &  $1 \pm 0$  & \;\, $1.21 \pm 0.2$  &  ---                    &   ---               \\
$64$     &   $201.122258$    &   $47.722542$    &  ILTJ132429.34+474321.1  &  5BZQJ1324+4743  &   ---               &   FSRQ                    & 2.26 &  $227 \pm 45$  &  $-0.09 \pm 0.14$  &  ---                    &   ---               \\
$65$     &   $201.854737$    &   $50.146967$    &  ILTJ132725.14+500849.0  &  5BZQJ1327+5008  &   ---               &   FSRQ                    & 1.025 &  $786 \pm 157$  &  $-0.43 \pm 0.12$  &  ---                    &   ---               \\
$66$     &   $202.520648$    &   $54.247706$    &  ILTJ133004.96+541451.7  &  5BZQJ1330+5414  &   ---               &   FSRQ                    & 0.839 &  $362 \pm 74$  &  $-0.65 \pm 0.14$  &  ---                    &   ---               \\
$67$     &   $202.677516$    &   $52.037599$    &  ILTJ133042.60+520215.3  &  5BZUJ1330+5202  &  3FGL J1330.9+5201  &   Uncertain               & 0.783 &  $8 \pm 2$  & \;\, $1.03 \pm 0.2$  & \num{7.24(151)e-14} &   $2.249 \pm 0.202$ \\
$68$     &   $203.188697$    &   $47.372615$    &  ILTJ133245.29+472221.4  &  5BZQJ1332+4722  &  3FGL J1331.8+4718  &   FSRQ                    & 0.669 &  $213 \pm 43$  & \;\, $0.09 \pm 0.17$  & \num{2.05(26)e-12} &   $2.529 \pm 0.104$ \\
$69$     &   $203.474375$    &   $50.959978$    &  ILTJ133353.85+505735.9  &  5BZUJ1333+5057  &  3FGL J1333.7+5057  &   Uncertain               & 1.362 &  $68 \pm 14$  &  $-0.16 \pm 0.15$  & \num{2.34(20)e-12} &   $2.496 \pm 0.071$ \\
$70$     &   $203.654959$    &   $56.529912$    &  ILTJ133437.19+563147.6  &  5BZQJ1334+5631  &   ---               &   FSRQ                    & 0.343 &  $714 \pm 143$  &  $-0.62 \pm 0.12$  &  ---                    &   ---               \\
$71$     &   $204.45688$     &   $55.017399$    &  ILTJ133749.65+550102.6  &  5BZQJ1337+5501  &   ---               &   FSRQ                    & 1.099 &  $586 \pm 117$  & \;\, $0.03 \pm 0.13$  &  ---                    &   ---               \\
$72$     &   $205.400755$    &   $55.243832$    &  ILTJ134136.18+551437.8  &  5BZBJ1341+5514  &  3FGL J1341.5+5517  &   BL Lac                  & 0.207 &  $54 \pm 11$  &  $-0.1 \pm 0.17$  & \num{6.12(112)e-13} &   $2.563 \pm 0.169$ \\
$73$     &   $206.439252$    &   $53.548524$    &  ILTJ134545.42+533254.6  &  5BZUJ1345+5332  &   ---               &   Uncertain               & 0.136 &  $1334 \pm 267$  &  $-0.5 \pm 0.15$  &  ---                    &   ---               \\
$74$     &   $207.394163$    &   $53.688067$    &  ILTJ134934.60+534117.0  &  5BZQJ1349+5341  &  ---                &   FSRQ                    & 0.979 &  $2233 \pm 447$  &  $-0.31 \pm 0.14$  &  ---                    &   ---               \\
$75$     &   $207.992576$    &   $55.702997$    &  ILTJ135158.22+554210.7  &  5BZBJ1351+5542  &  ---                &   BL Lac                  & 0.224 &  $35 \pm 7$  & \;\, $0.52 \pm 0.22$  &  ---                    &   ---               \\
$76$     &   $208.366892$    &   $56.015809$    &  ILTJ135328.05+560056.9  &  5BZBJ1353+5600  &  ---                &   BL Lac                  & 0.404 &  $17 \pm 3$  &  $-0.04 \pm 0.17$  &  ---                    &   ---               \\
$77$     &   $209.773803$    &   $55.741439$    &  ILTJ135905.71+554429.1  &  5BZQJ1359+5544  &  3FGL J1359.0+5544  &   FSRQ                    & 1.013 &  $96 \pm 19$  & \;\, $0.46 \pm 0.41$  & \num{5.89(42)e-12} &   $2.593 \pm 0.068$ \\
$78$     &   $213.903322$    &   $48.508372$    &  ILTJ141536.80+483030.1  &  5BZBJ1415+4830  &  3FGL J1415.2+4832  &   BL Lac                  & 0.496 &  $21 \pm 4$  & \;\, $0.24 \pm 0.3$  & \num{1.89(29)e-12} &   $2.669 \pm 0.155$ \\
$79$     &   $214.281187$    &   $46.11776$     &  ILTJ141707.48+460703.9  &  5BZQJ1417+4607  &  ---                &   FSRQ                    & 1.554 &  $2988 \pm 598$  &  $-0.5 \pm 0.13$  &  ---                    &   ---               \\
$80$     &   $214.939704$    &   $54.387341$    &  ILTJ141945.53+542314.4  &  5BZBJ1419+5423  &  3FGL J1419.9+5425  &   BL Lac                  & 0.152 &  $640 \pm 128$  & \;\, $0.09 \pm 0.17$  & \num{1.62(11)e-12} &   $2.308 \pm 0.062$ \\
$81$     &   $215.086649$    &   $46.412448$    &  ILTJ142020.80+462444.8  &  5BZQJ1420+4624  &  ---                &   FSRQ                    & 1.244 &  $6 \pm 1$  & \;\, $0.87 \pm 0.2$  &  ---                    &   ---               \\
$82$     &   $215.346061$    &   $46.763346$    &  ILTJ142123.05+464548.0  &  5BZQJ1421+4645  &  ---                &   FSRQ                    & 1.669 &  $38 \pm 8$  & \;\, $0.52 \pm 0.2$  &  ---                    &   ---               \\
$83$     &   $215.775784$    &   $48.036404$    &  ILTJ142306.19+480211.0  &  5BZQJ1423+4802  &  ---                &   FSRQ                    & 2.232 &  $397 \pm 79$  &  $-0.02 \pm 0.13$  &  ---                    &   ---               \\
$84$     &   $216.873141$    &   $54.155871$    &  ILTJ142729.55+540921.1  &  5BZGJ1427+5409  &  ---                &  Galaxy-dominated BL Lac  & 0.106 &  $128 \pm 26$  &  $-0.46 \pm 0.14$  &  ---                    &   ---               \\
$85$     &   $217.341103$    &   $54.103115$    &  ILTJ142921.86+540611.2  &  5BZQJ1429+5406  &  ---                &   FSRQ                    & 3.03 &  $3550 \pm 710$  &  $-0.58 \pm 0.13$  &  ---                    &   ---               \\
$86$     &   $219.239702$    &   $56.656725$    &  ILTJ143657.53+563924.2  &  5BZBJ1436+5639  &  3FGL J1436.8+5639  &   BL Lac                  & 0.756 &  $52 \pm 10$  &  $-0.39 \pm 0.17$  & \num{7.00(96)e-14} &   $1.985 \pm 0.127$ \\
$87$     &   $219.319563$    &   $47.29049$     &  ILTJ143716.70+471725.7  &  5BZBJ1437+4717  &  ---                &   BL Lac                  & 0.944 &  $468 \pm 94$  &  $-0.74 \pm 0.13$  &  ---                    &   ---               \\
$88$     &   $219.942255$    &   $46.053169$    &  ILTJ143946.14+460311.4  &  5BZBJ1439+4603  &  ---                &   BL Lac                  & 1.023 &  $185 \pm 37$  &  $-0.12 \pm 0.12$  &  ---                    &   ---               \\
$89$     &   $219.946582$    &   $49.968528$    &  ILTJ143947.18+495806.7  &  5BZBJ1439+4958  &  3FGL J1440.1+4955  &   BL Lac                  & 0.174 &  $218 \pm 46$  &  $-0.27 \pm 0.15$  & \num{3.45(37)e-12} &   $2.602 \pm 0.105$ \\
$90$     &   $220.756355$    &   $52.027279$    &  ILTJ144301.53+520138.2  &  ---             &  3FGL J1442.6+5156  &   Radio galaxy            & 0.141 &  $14098 \pm 2821$  &  $-0.78 \pm 0.12$  & \num{1.83(51)e-14} &   $1.912 \pm 0.184$ \\
$91$     &   $222.750341$    &   $52.019887$    &  ILTJ145100.08+520111.5  &  5BZBJ1450+5201  &  3FGL J1450.9+5200  &   BL Lac                  & 2.471 &  $69 \pm 14$  & \;\, $0.31 \pm 0.34$  & \num{2.41(33)e-13} &   $2.177 \pm 0.094$ \\
$92$     &   $223.614342$    &   $51.409562$    &  ILTJ145427.44+512434.4  &  5BZBJ1454+5124  &  3FGL J1454.5+5124  &   BL Lac                  & 1.083 &  $654 \pm 131$  &  $-0.58 \pm 0.13$  & \num{1.59(8)e-12} &   $2.084 \pm 0.036$ \\
$93$     &   $224.015284$    &   $50.807276$    &  ILTJ145603.67+504826.2  &  5BZBJ1456+5048  &  ---                &   BL Lac                  & 1.379 &  $30 \pm 6$  &  ---                    &  ---                    &   ---               \\
$94$     &   $224.614145$    &   $48.546108$    &  ILTJ145827.39+483245.9  &  5BZBJ1458+4832  &  ---                &   BL Lac                  & 0.541 &  $3 \pm 1$  & \;\, $0 \pm 0.2$  &  ---                    &   ---               \\
$95$     &   $225.203022$    &   $47.853805$    &  ILTJ150048.73+475113.7  &  5BZQJ1500+4751  &  3FGL J1500.6+4750  &   FSRQ                    & 1.059 &  $565 \pm 113$  &  $-0.18 \pm 0.14$  & \num{7.07(148)e-13} &   $2.643 \pm 0.149$ \\
$96$     &   $225.27563$     &   $55.464006$    &  ILTJ150106.15+552750.4  &  5BZBJ1501+5527  &  ---                &   BL Lac                  &  \;\;---  &  $9 \pm 2$  & \;\, $0.08 \pm 0.2$  &  ---                    &   ---               \\
$97$     &   $225.950377$    &   $47.992007$    &  ILTJ150348.09+475931.2  &  5BZBJ1503+4759  &  3FGL J1503.7+4759  &   BL Lac                  & 0.159 &  $69 \pm 14$  & \;\, $0.06 \pm 0.17$  & \num{1.26(19)e-13} &   $2.218 \pm 0.134$ \\
$98$     &   $226.683827$    &   $49.565557$    &  ILTJ150644.12+493356.0  &  5BZQJ1506+4933  &  ---                &   FSRQ                    & 1.544 &  $91 \pm 20$  &  $-0.04 \pm 0.18$  &   ---                   &   ---  \\* \bottomrule
\end{longtable}
\end{landscape}
\end{longtab}

\begin{acknowledgements}
SM acknowledges support from the Irish Research Council Postgraduate Scholarship and the Irish Research Council New Foundations Award.

RM gratefully acknowledges support from the European Research Council under the European Union's Seventh Framework Programme (FP/2007-2013) /ERC Advanced Grant RADIOLIFE-320745.

HR and KJD acknowledge support from the ERC Advanced Investigator programme NewClusters 321271.

LKM acknowledges support from Oxford Hintze Centre for Astrophysical Surveys, which is funded through generous support from the Hintze Family Charitable Foundation. This publication arises from research partly funded by the John Fell Oxford University Press (OUP) Research Fund.

PNB is grateful for support from the UK STFC via grant ST/M001229/1.

GG acknowledges the CSIRO OCE Postdoctoral Fellowship.

MJH acknowledges support from the UK Science and Technology Facilities Council [ST/M001008/1].

IP acknowledges support from INAF under PRIN SKA/CTA `FORECaST'.

JS is grateful for support from the UK STFC via grant ST/M001229/1.

WLW acknowledges support from the UK Science and Technology Facilities Council [ST/M001008/1].

This paper is based (in part) on data obtained with the International LOFAR Telescope (ILT) under project code s LC2\_038 and LC3\_008. LOFAR \citep{2013A&A...556A...2V} is the LOw Frequency ARray designed and constructed by ASTRON. It has observing, data processing, and data storage facilities in several countries, which are owned by various parties (each with their own funding sources), and are collectively operated by the ILT foundation under a joint scientific policy. The ILT resources have benefited from the following recent major funding sources: CNRS-INSU, Observatoire de Paris and Universit\`e d'Orl\`eans, France; BMBF, MIWF-NRW, MPG, Germany; Science Foundation Ireland (SFI), Department of Business, Enterprise and Innovation (DBEI), Ireland; NWO, The Netherlands; The Science and Technology Facilities Council, UK; Ministry of Science and Higher Education, Poland.

Part of this work was carried out on the Dutch national e-infrastructure with the support of the SURF Cooperative through grant e-infra 160022 \& 160152. The LOFAR software and dedicated reduction packages on \url{https://github.com/apmechev/GRID\_LRT} were deployed on the e-infrastructure by the LOFAR e-infragroup, consisting of J. B. R. Oonk (ASTRON \& Leiden Observatory), A. P. Mechev (Leiden Observatory) and T. Shimwell (ASTRON) with support from N. Danezi (SURFsara) and C. Schrijvers (SURFsara).

This research has made use of data analysed using the University of Hertfordshire high-performance computing facility (\url{http://uhhpc.herts.ac.uk/}) and the LOFAR-UK computing facility located at the University of Hertfordshire and supported by STFC
[ST/P000096/1].

This research has made use of the NASA/IPAC Extragalactic Database (NED), which is operated by the Jet Propulsion Laboratory, California Institute of Technology, under contract with the National Aeronautics and Space Administration.

This research has made use of SciPy software \citep{scipy} and Topcat \citep{2005ASPC..347...29T}.
\end{acknowledgements}
\bibliographystyle{aa}
\bibliography{bibliography}

\end{document}